\documentclass{aa}  

\usepackage{graphicx}
\usepackage{subfigure}
\usepackage{txfonts}
\usepackage{hyperref}
\usepackage{xcolor}

\usepackage{amsmath}
\usepackage{algorithm}
\usepackage[noend]{algpseudocode}
\usepackage{mathrsfs} 

\newcommand{\norm}[1]{\left\lVert#1\right\rVert}

\floatname{algorithm}{Procedure}

\graphicspath{{./}{figures/}}

\begin{document}

\title{DoG-HiT: A novel VLBI Multiscale Imaging Approach}

\author{H.Müller
        \inst{1}
        \and
        A.P. Lobanov\inst{1}
        }

\institute{Max-Planck-Institut für Radioastronomie,
        Auf dem Hügel 69, Bonn, 53121, Germany\\
        \email{hmueller@mpifr-bonn.mpg.de}, \email{alobanov@mpifr-bonn.mpg.de}
        }
        
\date{Received September 15, 1996; accepted March 16, 1997}

\abstract
{Reconstructing images from very long baseline interferometry (VLBI) data with sparse sampling of the Fourier domain ({\em uv}-coverage) constitutes an ill-posed deconvolution problem. It requires application of robust algorithms maximizing the information extraction from all of the sampled spatial scales and minimizing the influence of the unsampled scales on image quality.}
{We develop a new multiscale wavelet deconvolution algorithm DoG-HiT for imaging sparsely sampled interferometric data which combines the difference of Gaussian (DoG) wavelets and hard image thresholding (HiT). Based on DoG-HiT, we propose a multi-step imaging pipeline for analysis of interferometric data.}
{DoG-HiT applies the compressed sensing approach to imaging by employing a flexible DoG wavelet dictionary which is designed to adapt smoothly to the {\em uv}-coverage. It uses closure properties as data fidelity terms only initially and perform non-convex, non-smooth optimization by an amplitude conserving and total flux conserving hard thresholding splitting. DoG-HiT calculates a multiresolution support as a side product. The final reconstruction is refined through self-calibration loops and imaging with amplitude and phase information applied for the multiresolution support only.}
{We demonstrate the stability of DoG-HiT and benchmark its performance against image reconstructions made with CLEAN and Regularized Maximum-Likelihood (RML) methods using synthetic data. The comparison shows that DoG-HiT matches the superresolution achieved by the RML reconstructions and surpasses the sensitivity to extended emission reached by CLEAN.}
{Application of regularized maximum likelihood methods outfitted with flexible multiscale wavelet dictionaries to imaging of interferometric data matches the performance of state-of-the art convex optimization imaging algorithms and requires fewer prior and user defined constraints.} 


\maketitle

\section{Introduction} \label{sec:intro}
In very long baseline interferometry (VLBI), signals recorded at individual radio antennas are combined (correlated) in order to sample angular scales inversely proportional to pairwise antenna separations projected onto the plane of the incoming wavefront. Described by the van Cittert-Zernike theorem, the correlation product (visibility) of the signals recorded at two antennas at a given time is given by a single spectral harmonic corresponding to a single spatial frequency of the Fourier transform of the observed sky brightness distribution \cite[see][]{Thompson1994}. From a complete sampling of spatial frequencies, the {\em }true image could be revealed by the inverse Fourier transform. However, the  practical limitations on the number of antennas, observing bandwidth and observing time often result in situation when VLBI data provide only sparse sampling (uv-coverage) of the spatial frequencies (or "Fourier domain"), below the Nyquist-Shannon sampling rate.

The development of powerful imaging algorithms such as CLEAN \citep{Hogbom1974} and their decade long successful application in VLBI studies demonstrated that a reliable reconstruction of the true sky brightness distribution is still possible under a strong assumption about the sky brightness distribution being compressible as a sum of point sources. CLEAN and its many variants \citep[e.g.][]{Clark1980, Schwab1984} work well not only for compact structures but also for extended emission. CLEAN is still broadly used, mainly because it is practical. However, frontline VLBI applications such as millimetre or space VLBI demand better imaging tools which would alleviate the known limitations of CLEAN and provide superresolution, multiscalar decompositions, and high dynamic range.

Multiresolution imaging routines based on the greedy matching pursuit procedure inherent to CLEAN have been developed for decades now \citep{Wakker1988, Starck1994, Bhatnagar2004, Cornwell2008, Rau2011}. These studies build up on the great success of compressed sensing theory \citep[e.g.][]{Candes2006, Donoho2006}, i.e. that an image can be sparsely represented in a suitable set of basis functions (atoms). Even the CLEAN algorithm (sparsity in pixel basis) \citet{Lannes1997} and total variation regularization methods (sparsity of the Haar wavelet) could be understood in this way.

Imaging algorithms based on wavelets attract close attention of the astronomy community because they stand out as extremely helpful in the analysis and compression of image features on multiple scales \citep{Starck2006, Starck2015, Mertens2015, Line2020}. Both extended emission features and small scale structures are well compressible with wavelets. Moreover, wavelets of varying scales are sensitive to different ranges of visibilities, allowing the user to incorporate information about the radially distributed positions of gaps in the uv-coverage in the imaging procedure. Hence, sparsity in the wavelet domain is a strong and interesting image prior for the radio aperture synthesis imaging problem. 

The past five decades saw an ongoing development of regularized maximum likelihood methods (RML) for interferometric imaging, in particular with the development of image entropy regularizers such as the maximum entropy method (MEM)  \citep[e.g.][]{Frieden1972, Narayan1986, Wiaux2009, Li2011, Garsden2015, Thiebaut2017}. The RML methods have been applied particularly extensively for imaging with the EHT\footnote{Event Horizon Telescope}, \citep[e.g.][]{Ikeda2016, Akiyama2017, Akiyama2017b, Chael2018, eht2019d}. In a typical RML application, the image is recovered by minimizing simultaneously a data fidelity term which measures the proximity of the recovered solution to the true data, i.e visibilities and/or closure properties, and a set of regularization terms which measure the feasibility of the recovered solution. It has been demonstrated that $l_1$-penalty terms promote sparsity in the image domain. Hence, RML algorithms and the progress in convex optimization \citep{Beck2009, Combettes2009} provide a powerful framework of respecting sparsity during image deconvolution.

However, the deficiencies of uv-coverages inherent to such interferometric instruments as the EHT or the space VLBI mission {\em RadioAstron} pose additional challenges. Compressed sensing approaches applied to data from such arrays are capable, in principle, of recovering the significant structure of the target (achieving small data fidelity terms) while suppressing any additional noise-induced image features and sidelobes (achieving small penalty terms). However, in VLBI observations the sidelobes and the true image structure often become comparable in their magnitudes. The suppression of structure due to image sparsity affects the recovered data significantly. A more advanced treatment of image features, i.e. a more advanced differentiation between observed emission and noisy uv-gap induced structures, and an amplitude conserving optimization strategy are needed. Furthermore, an unsupervised approach for blind imaging is desired.

Random and systematic noise factors in the final image can be induced at various steps of the analysis. In particular, errors resulting from  uv-coverage deficiencies and antenna based noise factors (calibration issues, thermal noise) depend on location of the trace of the antenna pair in the uv-plane. Hence, these errors are scale and direction dependent. We need a novel algorithm that can deal with this, i.e. that can automatically decompose noisy features from signal features. This is a task that is suitable for wavelets in the first instance since they decompose the image into a sequence of scales. Direction dependent information is more difficult to compress and will not be addressed in this paper.

In this paper, we present a new multiscalar wavelet imaging algorithm built upon the compressed sensing approach. Our method extends over standard sparsity promoting imaging algorithms by applying a more stringent separation of significant image features from noise contributions by using an adaptive wavelet dictionary and suppressing the noise-induced artifacts in a novel amplitude-conserving hard thresholding algorithm. This algorithm is well suited for dealing with high level sidelobes such as the ones typically found in the data from EHT or space VLBI observations. 

An important feature of the algorithm is that the initial selection of the scales in the wavelet dictionary derives from the uv-coverage of observations and not from any assumptions about structure of the target source. We utilize current state of the art optimization algorithms for solving the resulting RML minimization problem, but amend the imaging pipeline by a hard thresholding sparsity term based on the multiresolution support, which allows us to retain necessary image information while suppressing noisy scales. We deal with potential residual calibration deficiencies of the data by first using only the gain-invariant closure quantities for imaging and then, after identifying and suppressing noise contributions, imaging the full data with an optimized, fixed multiresolution support. The resulting objective functional for minimization is not convex and not smooth, which requires employing non-convex and non-smooth optimization strategies. We present a final imaging pipeline which is immediately applicable to VLBI data. This imaging pipeline takes considerably fewer parameters than typical RML pipelines, thereby presenting a viable step towards a more unsupervised imaging approach. 

We test our pipeline routine on test images recently used to verify the modern generation of RML image routines \citep[][]{Tiede2020}. For incomplete uv-coverages, our algorithm performs better than the canonical CLEAN and its multiscale variants, owing to the flexibility of the dictionary (allowing to adapt it to a specific uv-coverage of the array), the sparse representation of astronomical images in the wavelet basis (compared to the representation with CLEAN or MS-CLEAN components), and the correct treatment of scale-dependent noise properties.

\section{Theory} \label{sec: the}

This section summarizes the relevant theory and background for different aspects of the new algorithm, focusing primarily on application of wavelets for deconvolution in aperture synthesis and on specific aspects of optimization procedures applied to sparsely sampled data. 

\subsection{Aperture Synthesis} \label{ssec: ape}

In interferometric observations, every antenna in the array records the electromagnetic field of an incoherent sky brightness distribution $I(x, y)$, where $x$ and $y$ are angular coordinates on the sky. Following the van Cittert-Zernike theorem, the cross correlation between the signals recorded by two antennas over a baseline $(u, v)$ (spatial frequencies in units of wavelengths) is given by the Fourier transform of $I(x, y)$ at this baseline: 
\begin{align}
    \mathcal{V} (u, v) = \int \int e^{-2 \pi i (x u + y v)} I(x, y) dx dy \,,  \label{eq: vis}
\end{align}
where $\mathcal{V}$ is the complex visibility. This relation holds under assumptions of a flat wavefront and small field of view approximation. Every antenna pair at a fixed time gives rise to a specific baseline. The projection of a baseline on a plane orthogonal to the direction to the target the Earth rotates baselines smoothly shift by time describing the typical elliptical traces in uv-coverages. However, due to the small number of antennas in VLBI arrays the coverage of measurements in the uv-domain remains sparse. In particular, gaps in the uv-coverage introduce sidelobes and artifacts in the recovered image. When inverting the Fourier transform (to produce $I^D$) the result can be written as convolution:
\begin{align}
    I^\mathrm{D} = B * I, \label{eq: dirty_map}
\end{align}
where $I^\mathrm{D}$ is the dirty image, i.e. the inverse Fourier transform of the (tapered) observed visibilities, and $B$ is the dirty beam, i.e. the inverse Fourier transform of the (tapered) projection onto measured baselines in the Fourier codomain.

Aperture synthesis imaging is the problem of recovering the true distribution $I(x, y)$ from a discrete sparse set of observed visibilities. This procedure could also be understood as a deconvolution problem, see Eq. \eqref{eq: dirty_map} . The incomplete uv-coverage introduces direction- and scale-dependent sidelobe patterns in the dirty image and the dirty beam. Deconvolution in this case becomes an ill-posed inverse problem. In particular, the solution to the imaging problem in Eq. \eqref{eq: dirty_map} is strictly speaking not unique as there are Fourier harmonics missing from the observation (historically called the ``invisible distributions'').  A successful deconvolution method must be able to identify and categorize these invisible distribution and minimize their impact on the restored image.

Image restoration is further complicated by the variable thermal noise and signal-to-noise ratio (SNR) of visibility measurements. The visibility SNR is systematically reduced at long baselines. As the antenna sensitivity enters the reconstruction at specific scales and directions, determined by the position of the baseline corresponding to a given antenna pair, the noise becomes scale and direction dependent. 

Various calibration issues also need to be addressed during image restoration. Systematic direction-independent calibration errors can be factorized into multiplicative station based gains $g_\mathrm{i}$ (where the index $i$ denotes the antenna in the array) affecting the relation between the observed visibilities $V_\mathrm{ij}$ and the true visibilities $\mathcal{V}_\mathrm{ij}$:  
\begin{align}
    V_\mathrm{ij} \approx g_\mathrm{i} g_\mathrm{j}^\star \mathcal{V}_\mathrm{ij} + N_\mathrm{ij}, \label{eq: obs_vis}
\end{align}
where $N_\mathrm{ij}$ denotes thermal noise on the baseline. In particular, phase information is typically only available after a calibration by an {\em ad hoc} initial model. In standard imaging approaches (e.g., in CLEAN), the problem of calibration is typically addressed through a hybrid imaging approach. In this case, an initial image is first produced using the a priori set of instrumental gains and then the gain terms are solved for as in Eq. \eqref{eq: obs_vis} in order to enforce consistency with the current image guess (with the solution typically obtained by a gradient descent approach or self-calibration), and these two steps are repeated iteratively until the desired image quality is reached. In this way alternating self-calibration and imaging steps converges to a self-contained model description consistent with the observed and self-calibrated data.

Some of the calibration issues can be circumvented by employing closure quantities computed from combinations of visibilities that are independent of antenna-based gain errors. The closure phase, $\Psi_\mathrm{ijk}$, is the phase over a triangle of antennas $\mathrm{i,j,k}$, i.e.:
\begin{align}
   \Psi_\mathrm{ijk} = \mathrm{arg} \left( V_\mathrm{ij} V_\mathrm{jk} V_\mathrm{ki} \right). 
\end{align}
The closure amplitude, $A$, is the ratio of amplitudes over a square of antennas $\mathrm{i,j,k,l}$:
\begin{align}
    A_\mathrm{ijkl} = \frac{|V_\mathrm{ij}| |V_\mathrm{kl}|}{|V_\mathrm{ik}| |V_\mathrm{jl}|}.
\end{align}
Not all closure triangles and closure squares are independent, which leads to reducing the number of total observables. Let us assume that at a specific time $N$ antennas are observing simultaneously. This gives rise to $N(N-1)/2$ independent baselines, while there are only $(N-1)(N-2)/2$ independent closure phases and $N(N-3)/2$ independent closure amplitudes \citep{Chael2018}. Hence the number of observables is reduced by a fraction of $1-2/N$ for closure phases and $1-2/(N-1)$ for closure amplitudes.

\subsection{Deconvolution}

Historically, the imaging problem described by Eq. \eqref{eq: dirty_map} has been addressed through inverse modelling, i.e. by CLEAN \citep{Hogbom1974} which can be classified as a greedy, matching pursuit algorithm. The problem is first translated into a deconvolution problem by taking the inverse Fourier transform of the visibilities. Hence, CLEAN requires performing this inversion on calibrated complex visibilities at every stage. The deconvolution problem is therefore solved by inverse modelling: CLEAN searches iteratively for the position of the maximum in the residual image, stores this in a list of delta-components, and updates the residual by subtracting the rescaled and shifted dirty beam from the residual image. In multiscale variants of CLEAN the delta components are replaced by more sophisticated extended basis functions \citep{Bhatnagar2004, Cornwell2008, Rau2011}. Recent years saw a continued development of imaging by forward modelling \citep[e.g.][]{Garsden2015, Akiyama2017, Chael2018} in which Eq. \eqref{eq: vis} is solved by fitting a model solution to the visibilities by minimizing the error in some cost functional (data fidelity term). With this forward modelling approach, RML methods can work directly on the closure quantities or a mix of data products in order to reduce the influence of calibration errors on the reconstruction. Regularization and missing information are dealt with by simultaneously minimizing a penalization term which promotes desired image features, i.e. sparsity, smoothness or small entropy. The resulting minimization problem is then solved by standard numerical optimization algorithms, e.g. by a gradient descent algorithm.

A major advantage of the work presented in this paper is the use of novel basis functions (i.e. wavelets). We will discuss them in more detail in Sec. \ref{ssec: wavelets}. These wavelets are extended and allow a more thorough analysis of the uv-coverage of the observations. The basis functions used in (MS-)CLEAN and RML are typically not offering this kind of analysis. Standard CLEAN \citep{Hogbom1974} models the image as a set of delta functions. It's multiscalar variants use some version of truncated Gaussian functions \citep[i.e. see the discussions in][]{Cornwell2008}. RML methods are utilizing pixel grids.

\subsection{Wavelets} \label{ssec: wavelets}
The continuous wavelet transform (CWT) could be understood as an extension of Fourier transform \citep{Starck2015} in which the Fourier decomposition in frequency domain is amended by a windowing of the measurement domain with a specially designed {\em analyzing wavelet} function. In the definition of \citet{Grossmann1989}, the CWT related to an analyzing wavelet $\Phi(t)$ operates in one dimension on the space of square integrable functions so that
\begin{align}
    I \mapsto W(a, b) = \frac{1}{\sqrt{a}} \int I(t) \Phi^* \left( \frac{t-b}{a} \right) dt = I * \tilde{\Phi}_{a}(b),
\end{align}
where $\tilde{\Phi}_a(t) = \frac{1}{\sqrt{a}} \Phi^*(\frac{-t}{a})$, $a$ is the scale parameter and $b$ is the position parameter. Hence, the CWT performs effectively a number of convolutions with dilated versions of the analyzing wavelet $\Phi$. There are different choices for the analyzing wavelet functions around including the Morlet wavelet \citep{Goupillaud1985, Coupinot1992}, the Haar wavelet \citep{Stollnitz1994}, the Mexican-hat wavelets \citep{Murenzi1989} and discrete versions \citep[e.g. see][and references therein]{Mallat1989, Starck2015}.

In this work we are using Difference of Gaussian (DoG) wavelets that are commonly applied to approximate Mexican-hat wavelets \citep{Gonzalez2006, Assirati2014}:
\begin{align} \nonumber
    \Phi_\mathrm{DoG}^{\sigma_1, \sigma_2}(x, y) &= \frac{1}{2 \pi \sigma_1^2} \exp \left( \frac{-r(x,y)^2}{2 \sigma_1^2} \right) - \frac{1}{2 \pi \sigma_2^2} \exp \left( \frac{-r(x,y)^2}{2 \sigma_2^2} \right) \\
    &= G_{\sigma_1} - G_{\sigma_2}, \label{eq: dog}
\end{align}
where necessarily $\sigma_1 \leq \sigma_2$ and $G_{\sigma_j}$ denotes a Gaussian with standard deviation $\sigma_j$.

Wavelets in image domain (convolution) directly translate to masks in Fourier domain (pointwise multiplication).
\begin{align}
    \mathcal{F} \Phi_\mathrm{DoG}^{\sigma_1, \sigma_2}(u,v) \propto \exp \left( -2 \pi^2 \sigma_1^2 q(u,v)^2  \right) - \exp \left( -2 \pi^2 \sigma_2^2 q(u,v)^2  \right), \label{eq: fourier_dog} 
\end{align}
where $q(u,v)$ denotes the radius in Fourier domain.

Of special interest for image compression is the discrete wavelet transform, in particular the a-trou wavelet transform (also called starlet transform). In a nutshell, the a-trou wavelet transform aims to compute a sequence of smoothing scales $c_j$ by convolving the image with a discretized smoothing kernel dilated by $2^j$ pixels, where $j$ labels the scale and ranges from 0 up to a final smoothing scale $J$. Wavelet scales are defined as the difference of two smoothing scales:
\begin{align}
    \omega_j = c_j - c_{j+1}
\end{align}
The last smoothing scale $c_J$ is added to the set of wavelet scales resulting in the set: $\left[ \omega_0, \omega_1, ..., \omega_{J-1}, c_J \right]$. This set decomposes the initial image into subbands $\omega_j$, each of them containing information on spatial scales from $2^{j} \rho$ to $2^{j+1} \rho$, where $\rho$ is the smallest scale in the image, i.e. the width of the smoothing kernel (which is often chosen to be close to the pixel scale). The set is complete in the sense that the image at the limiting resolution $c_0$ can be recovered by summing all scales:
\begin{align}
    c_0 = \sum_j \omega_j + c_J \label{eq: wavelet_sum}\,.
\end{align}

The a-trou wavelet transform has a wide range of applications and was successfully applied to radio interferometry before \citep[e.g.][]{Li2011, Garsden2015}. However, the a-trou wavelet decomposition by construction allows only for scales with the widths of $2^0, 2^1, 2^2, 2^3, ...$ pixels. In this study, we are interested in getting a more flexible selection of scales to adapt the scales to the uv-coverage in order to differentiate better between well and poorly constrained spatial scales.

Therefore, we propose to construct a continuous wavelet decomposition out of DoG wavelets in the same way as the a-trou wavelet transform was constructed out of a discretized smoothing kernel. We select an ascending sequence of widths $\sigma_0 \leq \sigma_1 \leq ... \leq \sigma_J$ and compute the smoothing scales $c_j$ by convolution with Gaussians with widths $\sigma_j$, i.e. $c_j = I * G_{\sigma_j}$. The wavelet scales $\omega_j$ are then set by 
\begin{align}
    \omega_j = c_j - c_{j+1} = I * \Phi^{\sigma_j, \sigma_{j+1}}_{DoG}\,,
\end{align}
which approximates sufficiently well the Mexican hat wavelet scales.

We call a set of basis functions in compressed sensing a {\em dictionary}, while the basis functions itself are called {\em atoms} of the dictionary. The term dictionary is also used for the linear mapping that evaluates a coefficient array of these atoms \citep{Starck2015}. The set of DoG wavelet functions $\Phi^{\sigma_j, \sigma_{j+1}}_{DoG}$ together with the last smoothing scale $G_{\sigma_J}$ builds a multiscalar dictionary $\Gamma$:
\begin{align}
    \Gamma: (I_0, I_1, I_2, ..., I_J) \mapsto \sum_{j=0}^{J-1} \Phi^{\sigma_j, \sigma_{j+1}}_{DoG} * I_j + G_{\sigma_J} * I_J. \label{eq: final_dict}
\end{align}
The atoms of the dictionary $\Gamma$ are the wavelets $\Phi^{\sigma_j, \sigma_{j+1}}_{DoG}$ and $G_{\sigma_J}$. By construction, see also Eq. \eqref{eq: wavelet_sum}, all atoms in the dictionary sum to $G_{\sigma_0}$, which is (given that $\sigma_0$ should be chosen very small, i.e. $G_{\sigma_0}$ is a delta peak at the pixel scale) indicating that the dictionary $\Gamma$ has full rank.  

Another crucial property of the dictionary $\Gamma$ is that the integral of the atoms $\Phi_\mathrm{DoG}^{\sigma_j, \sigma_{j+1}}$ is vanishing. Hence, only the final smoothing scale $G_{\sigma_J}$ transports total flux in the image.

The subbands $I_j$ hold the information of the image at a respective scale described by $\sigma_j$ and $\sigma_{j+1}$. We will denote the collection of subbands of an image $I$ by $\mathscr{I} = \{I_1, I_2, ..., I_J\}$ for the rest of the paper. However, even if $I = \Gamma (I_1, I_2, ..., I_J)$ holds, it is usually $I_j \neq \omega_j$ due to the non-orthogonality of the DoG wavelet functions. However, $\omega_j$ should provide a reasonable initial guess if one tries to find an array $\mathscr{I} = \{I_1, I_2, ..., I_j\}$ which satisfies $I = \Gamma \mathscr{I}$.

\subsection{Sparsity Promoting Regularization}
We apply sparsity promoting regularization in the generalized Tikhonov framework:
\begin{align}
    \hat{\mathscr{I}} \in \mathrm{argmin}_\mathscr{I} \left[ S(F \Gamma \mathscr{I}, V) + \alpha R(\mathscr{I}) \right], \label{eq: tikhonov}
\end{align}
where $S$ is the data fidelity term which measures the proximity between the recovered visibilities $F \Gamma \mathscr{I}$
and the observed visibility data, $V$. The term $F$ denotes mapping of the image intensity onto the visibilities, i.e. it computes a tapered and weighted projection of the Fourier transform of $x$ on a discrete and fixed sampling. The term $R$ denotes the regularization term which measures the feasibility of the guess $\mathscr{I}$. The parameter $\alpha$ controls the bias between both terms. The final recovered image solution is then:
\begin{align}
    \hat{I} = \Gamma \hat{\mathscr{I}}.
\end{align}

The data fidelity terms used for this paper are introduces as follows. Let $\mathscr{V} = F \Gamma \mathscr{I}$ denote the visibility data predicted from the current guess. We quantify the proximity between the predicted and measured visibilities by the effective $\chi^2$-distance between them,
\begin{align}
    S_\mathrm{vis} (\mathscr{V}, V) = \frac{1}{N_\mathrm{vis}} \sum_{i=1}^{N_\mathrm{vis}} \frac{|\mathscr{V}_i - V_i|^2}{\Sigma_i^2},
\end{align}
where $N_\mathrm{vis}$ is the number of visibilities and $\Sigma_i$ the estimated thermal noise of a given visibility. This $\chi^2$ corresponds directly to a log-Likelihood, given uncorrelated Gaussian thermal noise on the different baselines. In addition to this, we also use similar distances defined for three additional quantities. The between the measured and predicted visibility amplitudes,
\begin{align}
    S_\mathrm{amp} (\mathscr{V}, V) = \frac{1}{N_\mathrm{vis}} \sum_{i=1}^{N_\mathrm{vis}} \frac{(|\mathscr{V}_i| - |V_i|)^2}{\Sigma_i^2}\,.
\end{align}
The distance between the measured and predicted closure phases,
\begin{align}
    S_\mathrm{cph} (\mathscr{V}, V) = \frac{1}{N_\mathrm{cph}} \sum_{i=1}^{N_\mathrm{cph}} \frac{|\Psi_i(\mathscr{V}) - \Psi_i(V)|^2}{\Sigma_{\mathrm{cph},i}^2}\,, \label{eq: cph}
\end{align}
where $N_\mathrm{cph}$ is the number of closure phase combinations, $\Sigma_{\mathrm{cph}, i}$ the noise on a closure phase $\Psi_i(V)$, and $\Psi_i(\mathscr{V})$ denotes the respective closure phase computed from the array of predicted visibilities, $\mathscr{V}$. And finally, the distance between measured and predicted closure amplitudes,
\begin{align}
    S_\mathrm{cla} (\mathscr{V}, V) = \frac{1}{N_\mathrm{cla}} \sum_{i=1}^{N_\mathrm{cla}} \frac{|\ln A_i(\mathscr{V}) - \ln A_i(V)|^2}{\Sigma_{\mathrm{cla},i}^2}\,, \label{eq: lca}
\end{align}
with similar conventions as for the closure phases. We would like to note here that Eq.~\eqref{eq: cph} and Eq.~\eqref{eq: lca} are only approximate expressions for the correct log-likelihoods for closure products. \citep[e.g.][]{Blackburn2020, Arras2022}. These approximations and combinations of them are applied in Sec.~\ref{sec: test} for the analysis of test data.

It is known that sparsity is promoted by convex pseudonorm functionals as regularization terms \citep{Starck2015}, e.g. by a term of the form:
\begin{align}
    R_\mathrm{l_0}(\mathscr{I}) = \| \mathscr{I} \|_\mathrm{l_0} = \sum_{j=0}^J \sum_i w_j |I_j^i|^0, \label{eq: l0_pen}
\end{align}
with weights $w_j = \max{\Psi_\mathrm{DoG}^{\sigma_j, \sigma_{j+1}}}$ (see our discussion in Sec. \ref{ssec: outline}) and $i$ referring to the pixels in the subbands.

Another type of regularization terms used for this work are characteristic functions, incorporating a total flux $f$ constraint
\begin{align}
    R_\mathrm{flux}(I, f) = \begin{cases}
        0 & \mathrm{total\:flux\:of\:} I = f \\
        \infty & \mathrm{else},
    \end{cases}
\end{align}
or a multiresolution support, $M$, such that
\begin{align}
    R_\mathrm{mrs}(\mathscr{I}, M) = \begin{cases}
        0 & \mathscr{I} \neq 0 \mathrm{\:only\:in\:} M \\
        \infty & \mathrm{else}.    
    \end{cases}
\end{align}
The multiresolution support $M$ is a subdomain of the parameter space occupied by $\mathscr{I} = \{ I_1, I_2, ..., I_J \}$, and it comprises the coefficients in $\mathscr{I}$ that are allowed to be unequal to zero. In this sense, $R_\mathrm{mrs}$ could be understood as a compact flux constraint (i.e. all coefficients in the subbands $I_1, I_2, ..., I_J$ outside of a compact core region are constrained to zero), a multiscale constraint (i.e. all coefficients within one uncovered subband $I_j$ are set to zero) or a combination of both.

\subsection{Optimization} \label{ssec: optimization}
We use a flexible dictionary of DoG-wavelets and minimize Eq. \eqref{eq: tikhonov} directly with convex optimization algorithms. Generally, a gradient descent algorithm could be used for this task as long as the data fidelity term and the penalty term are smooth (i.e. possess a gradient). However, for sparsity promoting algorithms the penalty term is typically non-smooth, i.e. the $l_0$-norm is not differentiable. In numerical optimization, it is common practice to use the $l_1$-norm as a convex approximation to the non-convex $l_0$-functional stated above \citep[e.g.][]{Starck2015}. As the $l_1$-norm is also not smooth (prohibiting gradient descent algorithms from use), powerful optimization strategies were developed in numerical mathematics that typically outperform smooth approximations to the $l_1$-norm. Several of such optimization strategies have been recently applied to aperture synthesis as well \citep{Li2011, Carrillo2012, Carrillo2014, Garsden2015, Girard2015, Onose2016, Sardarabadi2016, Akiyama2017, Akiyama2017b, Onose2017, Cai2018a, Cai2018b, Chael2018, Pratley2018, eht2019d}. These algorithms depend on the proximal point operator instead of the gradient. However, in the present work we are addressing a slightly more advanced problem of maintaining sufficient contrast in the image, and hence we are interested in the $l_0$ functional instead of its common convex approximation $l_1$. Moreover, this will allow us to construct a multiresolution support later on. Relying on the overall success of proximal-point based algorithms in dealing with this kind of optimization problems, we nevertheless attempt to address our minimization problem by a proximal point based optimization.

In the following we describe basic properties of the proximal point operator. If $H$ is a proper, convex and lower semi-continuous functional on a Hilbert space $\mathbb{X}$, then the proximity operator of $H$ is defined as the mapping \citep{Moreau1962}:
\begin{align}
\mathrm{prox}_{\tau, H} (z) = \mathrm{argmin}_{s \in \mathbb{X}} \left \{ H(s) + \frac{1}{2 \tau} \lVert s - z \rVert_\mathbb{X} \right \}, \label{eq: prox1}
\end{align}
and $\mathrm{prox}_{\tau, H}$ is well defined (i.e. there is a unique single-valued minimum). For a convex, proper and lower semi-continuous objective functional, such as the right hand side of Eq. \eqref{eq: prox1}, the zero element is in the subdifferential of the functional at the point of the minimum. Hence, $\hat{s} := \mathrm{prox}_{\tau, H} (z)$ satisfies:
\begin{align}
z - \hat{s} \in \tau \partial H[\hat{s}]. \label{eq: prox2}
\end{align}
The power of proximal operators comes from their fixed-point property. It follows directly from Eq. \eqref{eq: prox1} and Eq. \eqref{eq: prox2}, that:
\begin{align}
\hat{s} \in \mathrm{argmin}_s H(s) \iff \hat{s} = \mathrm{prox}_{\tau, H} (\hat{s})
\end{align}
independently of $\tau \geq 0$, for a sketch of the proof see Appendix \ref{sec: fixed_point}.

Hence, we can solve the minimization in Eq.~\eqref{eq: tikhonov} by fixed-point iterating the proximity operator. This procedure is exact in the sense that convergence proofs are available \citep[e.g.][]{Martinet1972}. For a combination of a smooth term (data-fidelity term) and a non-smooth term (penalty term) one ends up at a two step splitting minimization strategy consisting of a gradient descent step for the data fidelity term and one proximity step for the penalty term \citep{Combettes2009}. The forward-backward splitting algorithm is outlined in its general framework in Table~\ref{alg: splitting}. The two-step splitting is realized during the last step of the algorithm, when the current guess is updated by a proximal step and a gradient descent step.

Interestingly, despite being derived in the context of convex optimization, there are also local convergence proofs available for the case when $S$ and $R$ are not convex, but the penalty term remains lower semicontinuous, proper and satisfies the technical Kurdyka-Łojasiewicz property, e.g. see \citet{Attouch2013, Ochs2014, Xiao2015, Bot2016, Liang2016} or \citet{Bao2016} for a connection to wavelets. This is of special interest for radio aperture synthesis as the data fidelity terms $S_\mathrm{amp}$, $S_\mathrm{cph}$ and $S_\mathrm{cla}$ are indeed not convex. Local convergence to a steady point is known and under some circumstances even global convergence could be proven \citep[compare the discussion in][]{Liang2016}. Application in practice shows that, given a reasonable initial guess, local minima could be avoided \citep{Starck2015}.

One may wonder, if the difficulty with convex non-smooth penalty-functionals is now just transported to the probably troublesome minimization problem in the definition of the proximity operator in Eq. \eqref{eq: prox1}. But the proximity operator is known for a large number of examples and the computation is often not more time consumptive than one Landweber iteration. For example for the $l_0$-functional the proximity operator is \citep[e.g.][]{Starck2015}:
\begin{align}
    \mathrm{prox}_{\tau, \norm{\cdot}_{l_0}}(z) = \begin{cases}
    z & | z | > \sqrt{2 \tau} \\
    \mathrm{sign}(z) [0, z] & | z | = 0\\
    0 &  | z | < \sqrt{2 \tau} 
    \end{cases}, \label{eq: l0_prox}
\end{align}
where signum and absolute value are meant to be evaluated pointwise. This not always single valued since the $l_0$ norm is not convex. The proximal point operator of characteristic functions is the projection on the support of the characteristic function: i.e. in the case of the multiresolution support the function that nullifies all coefficients outside the multiresolution support, and (in the case of the total flux) the function that projects the current guess to the guess with the correct total flux.

\begin{table}
\caption{Forward-Backward Splitting for the minimization of $S+R$}

\begin{tabular}{p{0.45\textwidth}}
\hline \\
\end{tabular}

\begin{algorithmic}

\Require $S, R: \mathbb{X} \mapsto \mathbb{R}$ (convex)
\Require grad $S$ is $L$-Lipschitz continuous

\State Step size: $\tau \in (0,2/L)$, typical choice: $\tau = 1/\norm{grad(S)}^2$
\State Initial guess: $x_0 \in \mathbb{X}$

\While{$i = 0, 1, 2, ...$}

\State $x_{i+1} = \mathrm{prox}_{\tau, R} (x_i - \tau \mathrm{grad} S (x_i))$

\EndWhile \\

\end{algorithmic}

\begin{tabular}{p{0.45\textwidth}}
\hline \\
\end{tabular}

\label{alg: splitting}
\end{table}

\section{Pipeline}
We use the same notations as in the former subsections: $V$ are the observed visibilities, $f$ the prior compact total flux, $\Gamma$ the dictionary of composed of DoG wavelets, and $F$ the linear mapping of the image intensity to the tapered visibilities.

\subsection{Outline} \label{ssec: outline}
The core of our imaging method concerns solving the following optimization problem:
\begin{align} \nonumber
    \hat{\mathscr{I}} \in \mathrm{argmin}_\mathscr{I} &\left[  S_\mathrm{cph}(F \Gamma \mathscr{I}, V) + S_\mathrm{cla}(F \Gamma \mathscr{I}, V) \right. \\ 
    &\left. + \alpha \cdot R_\mathrm{l_0} (\mathscr{I}) + R_\mathrm{flux} (\mathscr{I}, f) \right], \label{eq: second_round_problem}
\end{align}
where we choose the maximum of the corresponding DoG wavelet function as weights $\omega_i$. We have only one regularization parameter $\alpha$ that controls the amount of suppression by hard thresholding. We like to emphasize the main motivations behind this optimization problem:
\begin{itemize}
    \item We use the more flexible DoG dictionary here, see Eq. \eqref{eq: final_dict}. This allows us to adapt the dictionary to the uv-coverage by separating scales that are well covered by observations from those are less accurate constrained by observations. This will allows us to better suppress the signal from the latter one.
    \item We initially use the closure properties as data fidelity term as a measure to reduce the effect of possible antenna-based calibration errors. \citet{Chael2018} demonstrated that this information is sufficient to recover the image when using strong regularization priors. In later imaging rounds, i.e. after several self-calibration steps, we are also starting to include amplitude and phase information.
    \item We use hard thresholding ($l_0$ pseudonorm regularization). This promotes sparsity. In the few works addressing multiscalar imaging for radio aperture synthesis \citep{Li2011, Carrillo2012, Carrillo2014, Garsden2015, Onose2016, Sardarabadi2016, Onose2017, Pratley2018} often the $l_1$-norm is used as a convex approximation to $R_\mathrm{l_0}$. This is standard for sparsity promoting inverse problems \citep[e.g.][]{Starck2015}. However, the $l_1$-norm suppresses both, image features and noisy structures. As it is important to preserve the amplitude on the well covered scales, we resort to using the non-convex $l_0$-pseudonorm as penalization. We weight the $l_0$ pseudonorms by the maximal peak of corresponding DoG wavelet basis function. This is done to avoid that the scale selection would have a strong effect on the choice of the best regularization parameter. In principle these weighting parameters could be considered as free regularization parameters as well. However, to meet our requirement of constructing an algorithm that is as unbiased and data-driven as possible, we restrict them in this work to the choice that seems most reasonable.
    \item It should be noted that $S_\mathrm{cph}$, $S_\mathrm{cla}$ and $R_\mathrm{l_0}$ are invariant against rescaling the coefficients $x$ (atoms) by a scale factor $\lambda \in \mathbb{R}$. To select the most feasible solution along this line, we select the one that matches the prior compact total flux.  
    \item There are more possible regularization terms available, for example the total variation or the total squared variation terms that are applied for the EHT imaging \citep{eht2019d}. However, finding suitable weighting parameters for the different data terms and penalty terms is somewhat unintuitive for such different types of regularizations. This task often requires large parameter surveys with feasible synthetic data. We aim to find a largely unsupervised algorithm with only a few free parameters.
\end{itemize}

Our optimization problem differs significantly from previous multiscalar RML imaging approaches \citep[e.g.][]{Li2011, Carrillo2012, Carrillo2014, Garsden2015, Girard2015, Onose2016, Sardarabadi2016, Onose2017, Cai2018a, Cai2018b, Pratley2018}. They used the starlet transform as dictionary (replaced by DoG dictionary), the distance of observed and predicted visibilities as data fidelity term (replaced by closure properties), and $l_1$ penalty terms (replaced by $l_0$ penalization). 

Nevertheless, our algorithm shares some similarities with RML reconstructions. The unpenalized minimization of the data fidelity terms would yield a high resolving reconstruction which fits the observed data points with a (too) high fidelity, but provides clearly unphysical highly oscillating fits of the visibilities in the gaps of the uv-coverage. Total variation and total squared variation penalization effectively smooth the recovered model to a reasonable extent, where the amount of smoothing is controlled by the trade-off between data fidelity term and penalization term. We achieve a similar effect by modeling the brightness density distribution with (as few as possible) smooth, extended basis functions.

\subsection{Pipeline} \label{ssec: pipeline}
The data fidelity terms $S_\mathrm{cph}$, $S_\mathrm{cla}$ and the regularization term $R_\mathrm{l_0}$ are not convex. Hence, the minimization problem stated in Eq. \eqref{eq: second_round_problem} strictly speaking may not have a single-valued minimum. Therefore, a careful imaging pipeline helping global convergence is needed. Note also that the representation of the image in wavelet scales is an overcomplete representation. Due to the resulting large arrays, computation could be slow. Computation time can in principle be reduced when starting from a reasonable initial guess instead of a flat image or a Gaussian prior.

On the other hand, the solution of Eq. \eqref{eq: second_round_problem} returns an adequate calibration image $\hat{I} = \Gamma \hat{x}$ and computes, on the fly, a multiresolution support $M$ (all the pixels that are unequal to zero). Further imaging rounds, including the self-calibrated visibilities, allow the solution only to vary in the multiresolution support and hence could sharpen the image further while respecting the sparsity assumption due to the multiresolution support. This approach is realized within the following imaging pipeline:

\begin{enumerate}
    \item \textbf{Single Scalar flux constraining imaging}: \\
    We minimize the term: 
    \begin{align} \nonumber
        \hat{I}_1 \in \mathrm{argmin}_I &S_\mathrm{amp}(FI, V) + S_\mathrm{cph}(FI, V) \\
        &+ S_\mathrm{cla}(FI, V) + R_\mathrm{flux}(I, f),
    \end{align}
    where no dictionary is involved. We do this by the fast minimization method available in the scipy package. In fact this imaging round is similar to the first imaging round with the ehtim imaging package for \citet{eht2019d}. This imaging round is used for finding a reasonable initial guess in order to reduce the overall computation time. We convolve the result with the instrument clean beam (to avoid local minima) and only use a few iterations, i.e. an incomplete decomposition. Finally, we have to find some wavelet coefficient array $\hat{\mathscr{I}}^1$ that satisfies $\hat{I}_1 = \Gamma \hat{\mathscr{I}}^1$. To satisfy Eq.~\eqref{eq: wavelet_sum}, we copy the intensity $\hat{I}_1$ in every scale $\hat{\mathscr{I}}^1_j = \hat{I}_1$, where $j$ denotes the scale in use.
    \item \textbf{Multiscalar closure property Hard Thresholding imaging}: \\
    This imaging round is the heart of the new algorithm. We solve Eq.~\eqref{eq: second_round_problem} by a forward-backward splitting approach.
    
    We start from the initial guess $\hat{\mathscr{I}}^1$ computed in the first imaging round and compute a scale discrete guess in order to minimize Eq. \eqref{eq: second_round_problem}. We start from the largest scales only (set all other subbands to zero), successively add smaller scales and larger thresholds. We stop at the scale at which the functional \eqref{eq: second_round_problem} is minimal, i.e. at the smoothing when accuracy of the fit and sparsity penalization balance. Lastly, we reestimate the thresholds for each scale individually starting from the smallest scales.
    
    We then minimize, starting from this initial guess, the functional with a forward-backward splitting strategy. We will explain this forward-backward splitting minimization strategy in Sec. \ref{ssec: minimization_algorithm}. An outline of the round 2 imaging algorithm is presented in Tab. \ref{alg: sparsity}.
    \item \textbf{Multiresolution imaging with visibility amplitudes}: \\
    We self-calibrate the data with the image guess derived in the second image round. Moreover, we compute the multiresolution support $M$ from the result $\hat{\mathscr{I}}^2$ of the second imaging round, i.e. we choose all non-zero elements of the multiscalar coefficient array $\hat{\mathscr{I}}^2$ as multiresolution support. We now solve the problem:
    \begin{align} \nonumber
        \hat{\mathscr{I}}^2 \in \mathrm{argmin}_\mathscr{I} &S_\mathrm{amp}(F\Gamma \mathscr{I}, V) + S_\mathrm{cph}(F\Gamma \mathscr{I}, V) \\
        &+ S_\mathrm{cla}(F\Gamma \mathscr{I}, V) + R_\mathrm{mrs}(\mathscr{I}, M).        
    \end{align}
    This is solved by a simple gradient descent algorithm starting from the initial guess $\hat{\mathscr{I}}^3$ in which only the gradient with respect to the coefficient in the multiresolution support is computed.
    \item \textbf{Multiresolution imaging with full visibilities}: \\
    After another self-calibration step, we solve the imaging problem:
    \begin{align}
        \hat{\mathscr{I}}^4  \in \mathrm{argmin}_\mathscr{I} S_\mathrm{vis} (F\Gamma \mathscr{I}, V) + R_\mathrm{mrs}(\mathscr{I}, M),
    \end{align}
    by a gradient descent algorithm only varying coefficients in the multiresolution support analog to the third imaging round.
    \item \textbf{Single Scalar visibility imaging}:
    Finally, we set all pixels with negative flux to zero flux and increase the match to the observed visibilities by a gradient descent algorithm minimizing $S_\mathrm{vis}(FI, V)$ in the pixel scale starting from $\hat{I}^5 = \Gamma \hat{\mathscr{4}}^5$. 
\end{enumerate}
The last three imaging rounds (in particular round 5) are optional and only refine the reconstruction. This will be discussed in our demonstration on synthetic data in Sec.~\ref{sec: test}.

\subsection{Minimization Algorithm} \label{ssec: minimization_algorithm}
We now discuss the minimization algorithm used to minimize Eq.~\eqref{eq: second_round_problem}. All other imaging rounds are based on smooth gradient descent imaging algorithms (rounds 3-5) or a smooth Newton type minimization (round 1). But Eq. \eqref{eq: second_round_problem} is neither convex nor smooth. However, the data fidelity terms are smooth with Lipschitz continuous derivatives and the $l_0$ pseudonorm is proper, lower-semicontinuous and satisfies the Kurdyka-Lojasiewicz property \citep[e.g.][]{Liang2016}. Thus, the Forward-Backward Splitting algorithm \ref{alg: splitting} remains applicable, see our discussion in Sec. \ref{ssec: optimization}. Additionally, recall that $S_\mathrm{cph}$, $S_\mathrm{cla}$ and $R_\mathrm{l_0}$ are invariant against rescaling the coefficient array by a scalar factor $\lambda$. We therefore propose the following iterative scheme:

We first minimize $ S_\mathrm{cph}(F \Gamma x, V) + S_\mathrm{cla}(F \Gamma x, V) + \alpha \cdot R_\mathrm{l_0} (x)$ by a fixed number of forward-backward splitting iterations, then we rescale the coefficient array by a scale factor such that $I = \Gamma x$ has a total flux matching the prior compact flux (letting the data fidelity terms and regularization terms unaffected), then we proceed with our forward-backward splitting algorithm, doing rescaling again and so on. The complete procedure is outlined in Tab. \ref{alg: sparsity}.

The needed proximal operator for the $l_0$ pseudonorm is computed in Eq. \eqref{eq: l0_prox}.

\begin{table*}
\caption{Wavelet Forward-Backward-Splitting: Pipeline round 2}

\begin{tabular}{p{0.45\textwidth}}
\hline \\
\end{tabular}

\begin{algorithmic}
\Require Visibilities: $V$
\Require Stepsize: $\tau$ (chosen artificially, such that algorithm converges)
\Require Regularization Parameter: $\alpha$
\Require Total flux: $f$ \\

\Comment{\textit{Precompute needed data terms and operators}}

\State Define a dictionary of basis functions(wavelets): $\Gamma$
\State Define a forward operator: $G: \mathscr{I} \mapsto \mathcal{F} \Gamma \mathscr{I}$ (note $G$ is linear)
\State Define a data-fidelity functional: $df: \mathscr{I} \mapsto S_\mathrm{lca}(V, G\mathscr{I})+S_\mathrm{cph}(V, \Gamma \mathscr{I})$
\State Precompute gradient of data-fidelity functional: $df^\prime[\mathscr{I}]$
\State Define a penalty term: $pen: \mathscr{I} \mapsto |support(x)|$ ($l_0$-norm)
\State Precompute proximal operator of penalty term: $prox_{\tau}$ (hard thrinkage operator, Eq. \eqref{eq: l0_prox})
\State $\mathscr{I} = \mathrm{initial guess}$ \\

\Comment{\textit{Find initial image thresholding by minimizing Eq. \eqref{eq: second_round_problem} on a predefined grid of thresholds}}

\State Define grid of possible thresholds: $t_i$
\For{$i=1,2,3, ...$}
\State Hard thresholding: $test_i = prox_{t_i}(\mathscr{I})$
\State $min_i = df(test_i)+\alpha pen(test_i)$
\EndFor
\State Find minimum $i$ and update initial guess $\mathscr{I} = prox_{t_i}(\mathscr{I})$
\State $min_{tot} = min_{i}$
\For{$j=0,1,2,...,J$}
\For{$i=1,2,3, ...$}
\State Hard thresholding single scale: $test_{i,j} = \{\mathscr{I}_1, ..., prox_{t_i}(\mathscr{I}_j), ..., \mathscr{I}_J \}$
\State $min_{i,j} = df(test_{i,j})+\alpha pen(test_{i,j})$
\If{$min_{i,j} < min_{tot}$}
\State $min_{tot} = min_{i,j}$
\State $\mathscr{I}_j = prox_{t_i}(\mathscr{I}_j)$
\EndIf
\EndFor
\EndFor \\

\Comment{\textit{Start forward-backward iterations from this guess}}

\While{stopping-rule 1}
\While{stopping-rule 2}
\State $\mathscr{I} = \mathscr{I} - \tau \cdot df^\prime[\mathscr{I}]$
\State $\mathscr{I} = prox_{\tau \cdot \alpha} (\mathscr{I})$    
\EndWhile \\
\State $\mathscr{I} = \mathscr{I} \cdot f/sum(\Gamma \mathscr{I})$
\EndWhile \\
\State Compute Multiresolution support $M = \{ \mathscr{I} \neq 0 \}$ \\
\Ensure $\mathscr{I}$ is aprroximate minimizer to Eq. \eqref{eq: second_round_problem}
\Ensure $\hat{I} = \Gamma \mathscr{I}$ is an approximation to the true sky brightness distribution
\Ensure As a byproduct $M$ is a reasonable multi-resolution support

\end{algorithmic}

\begin{tabular}{p{0.45\textwidth}}
\hline \\
\end{tabular}

\label{alg: sparsity}
\end{table*}

Iterative reweighted $l_1$-regularization proposed by \citet{Candes2008} provides an alternative approach to solve optimization problems with non-convex $l_0$-terms and is more common than our forward backward scheme. However, our rescaling approach to match the total flux would affect the reweighting step of the reweighted $l_1$-regularization method. So, it would introduce an additional layer of complexity in solving the optimization problem. This would fail our requirement of a preferably simple imaging algorithm with a small number of parameters to specify. 

\subsection{Selection of scales} \label{ssec: selection_scales}
Our DoG-wavelet dictionary is flexible in the sense that the Gaussian widths could be chosen to adapt to the uv-coverage. Hence, the selection of scales is data driven (e.g. by the uv-coverage) and should be performed automatically. We discuss in this section the automatic scale-width selection and outline the key points of this approach.

The Fourier transform of a two-dimensional DoG wavelet is a ring shaped mask, recall Eq. \eqref{eq: fourier_dog}. It is reasonable to select the masks such that well covered regions of the uv-space and poorly covered regions are separated. However, for very sparse arrays, there are no really well covered scales. In this situation, our selection should be also driven by the assertion that all the data points belonging to the same antenna pair should be covered in one scale. 

\begin{figure}
    \centering
    \includegraphics[width=0.5\textwidth]{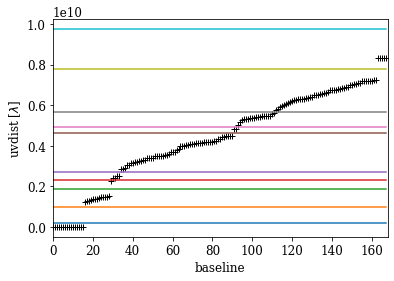}
    \caption{Sketch od automatic scale selection. The sorted array of uv-distances is plotted with black points. This array has clearly visible jumps (gaps in radial uv-coverage). We identify these jumps and assign scalar widths to it (colored horizontal lines).}
    \label{fig: scale_selection}
\end{figure}

We present a sketch of our automatic scale selection in Fig. \ref{fig: scale_selection}. We unpack the array of uv-distances of the full array, sort it in increasing order (black dots in Fig. \ref{fig: scale_selection}) and search for jumps between two consecutive data points that exceed a certain threshold. These jumps clearly appear at gaps in the uv-coverage (most visible between the blue and orange line in Fig. \ref{fig: scale_selection}, respectively between the green and the red line). We store the uv-distances at which these gaps appear and select the DoG wavelet widths by the mean of consecutive distances (represented by colored horizontal lines in Fig. \ref{fig: scale_selection}).

\begin{figure*}
    \centering
    \includegraphics[width=\textwidth]{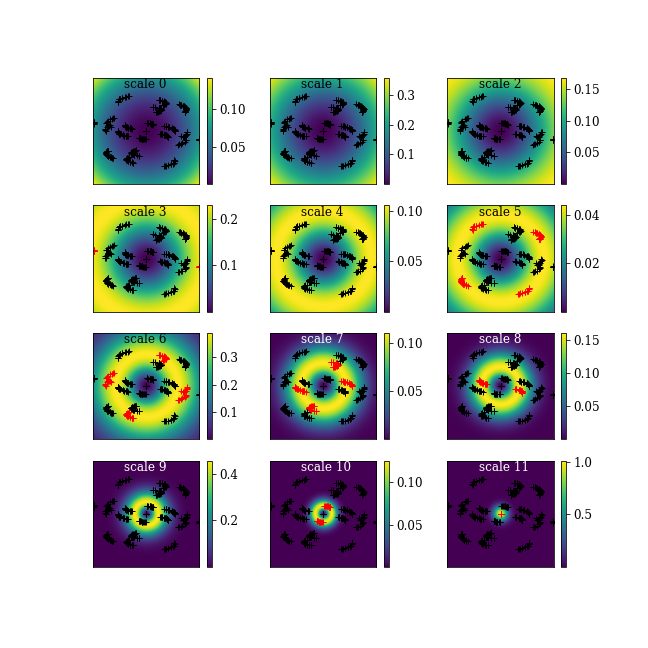}
    \caption{Observed uv-coverage (black/red points) of the EHT data array (observation of M\,87 on 5 April 2017) and masks defined by the DoG wavelets listed in Tab. \ref{tab: scales} (color maps). The masks are the Fourier transform of the respective wavelets and they define ring-like filters in Fourier domain. The visibilities highlighted by a specific filter are plotted in red}.
    \label{fig: all_scales}
\end{figure*}

As a demonstration, we apply this procedure to the EHT 2017 array. In Fig.~\ref{fig: all_scales}, we show our masks and the data points in uv-space. The widths information of the scales shown in Fig.~\ref{fig: all_scales} is given in Tab. \ref{tab: scales}. We also mention in Tab. \ref{tab: scales} which scale is most sensitive to which antenna pair, i.e. what was the selection criterion to this scale. As all DoG-wavelets satisfy the zero integral property of wavelets, then the only flux-transporting scale is the smoothing scale $G_{\sigma_J}$.

The smallest scale in our set has a width of $9.96\,\mu\mathrm{as}$ which corresponds to $5.02$ pixels in our discretization. For the sake of completing our dictionary of wavelet functions so that Eq. \eqref{eq: wavelet_sum} remains satisfied, we complete our sets of scales down to the pixel size by adding DoG wavelets according to the widths of 1, 2 and 4 pixels. This, however, will turn out to be less relevant as these scales will be suppressed by the algorithm automatically, see Sec. \ref{sec: imaging_pipeline}.

\begin{table}[]
    \centering
    \begin{tabular}{c|c|c}
        Scale & $\sigma_1 \rightarrow \sigma2$ ($\mu\mathrm{as}$) &  Main Sensitivity\\
         0&$0.84 \rightarrow 1.69$& Unresolved\\
         1&$1.69 \rightarrow 3.37$& Unresolved\\
         2&$3.37 \rightarrow 4.23$& Unresolved\\
         3&$4.23 \rightarrow 5.78$& PV-SMA/JCMT\\
         4&$5.78 \rightarrow 6.66$& Gap\\
         5&$6.66 \rightarrow 7.06$& AA/AP-JCMT/SMA, AA/AP-PV\\
         6&$7.06 \rightarrow 12.18$& AA/AP-SMT, SMT-PV, LMT-PV\\
         7&$12.18 \rightarrow 14.13$& AA/AP-LMT, JCMT/SMA-LMT\\
         8&$14.13 \rightarrow 17.55$& JCMT/SMA-SMT\\
         9&$17.55 \rightarrow 33.69$& Gap\\
         10&$33.69 \rightarrow 39.81$ & LMT-SMT \\
         11&$39.81$ & AA-AP, JCMT-SMA
    \end{tabular}
    \caption{Widths of DoG wavelets and their main sensitivity to the uv-coverage, i.e. which antenna pair is mainly covered by these scales. Differenet scales are most sensitive either to specific baselines or the gaps in the {\em uv}-coverage. The three smallest scales were added to complete the dictionary down to the pixel size and compress unresolved structures.}
    \label{tab: scales}
\end{table}

\section{Tests with synthetic data} \label{sec: test}

\subsection{Testdata} \label{ssec: testdata}
We test our algorithm on the same set of synthetic data that were recently used for testing feature extraction from the EHT data \citep{Tiede2020}. In particular, we use a crescent, a disk, a double Gaussian, and a ring structure.

The crescent is described by the Equation \citep{Tiede2020}: 
\begin{align}
    I(r, \theta) = I_0 (1-s \cos( \theta - \xi)) \frac{\delta(r-r_0)}{2 \pi r_0}.
\end{align}
We use $\xi = 180^\circ$, $r_0 = 22\,\mathrm{\mu as}$, $s=0.46$ and $I_0 = 0.6\,\mathrm{Jy}$. The crescent is then convolved with a Gaussian with the full width at half maximum (FWHM) of $10\,\mathrm{\mu as}$.

The disk is a disk of diameter $70\,\mathrm{\mu as}$. The disk is then convolved with a Gaussian with FWHM $10\,\mathrm{\mu as}$.

The double Gaussian image consists of two Gaussian peaks of FWHM $20\,\mathrm{\mu as}$. The first Gaussian is placed at the origin and has a flux of $0.27\,\mathrm{Jy}$. The second Gaussian is placed $30\,\mathrm{\mu as}$ to the East and $12\,\mathrm{\mu as}$ to the South. It has a flux of $0.33\,\mathrm{Jy}$.

The ring has radius of $22\,\mathrm{\mu as}$ and a total flux of $0.6\,\mathrm{Jy}$. The ring is convolved with a Gaussian with FWHM of $10\,\mathrm{\mu as}$.

We simulate visibility data from the test images with the help of the ehtim package, using the EHT 2017 array at $229\,\mathrm{GHz}$. We mimic the observation with the \textit{observe\_same} option assuming the same systematic noise levels, observation intervals and correlation times as for the EHT observations \citep{eht2019d}. We assume phase and gain calibration, but add thermal noise. 

We aim to study the image on a 128x128 pixel grid with $1\mu\mathrm{as}$-pixels. However, to avoid boundary effects in the computation (the largest chosen Gaussian has a FWHM of already $93.75\,\mu\mathrm{as}$), we widen the field of view by a factor of two. Moreover, we use $129$ pixels instead of $128$ pixels to discretize narrow central Gaussians correctly. We have defined $12$ different wavelet scales. Thus, we are attempting to solve for $12 \cdot 129 \cdot 129 \approx 2 \cdot 10^5$ parameters in the multiscale imaging rounds.

\subsection{Imaging Pipeline} \label{sec: imaging_pipeline}
In this subsection, we use with the crescent image to demonstrate the stability of our imaging pipeline and present some key features. 

\begin{figure*}
    \centering
    \includegraphics[width=\textwidth]{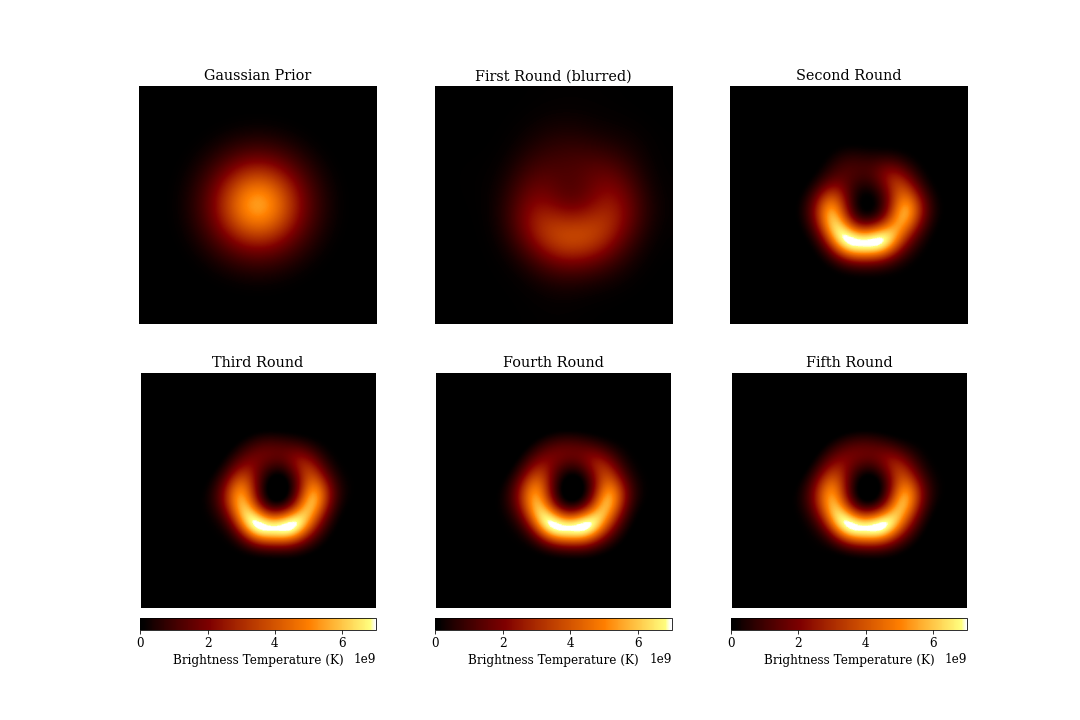}
    \caption{Imaging results of the crescent at various steps of the imaging pipeline. Upper left: Gaussian prior image. Upper middle: Initial guess, result after round 1 blurred by the $20\,\mu\mathrm{as}$ beam. Upper right: After imaging round 2. Bottom left: After Imaging round 3. Bottom middle: After imaging round 4. Bottom right: Final image after imaging round 5.}
    \label{fig: pipeline_images}
\end{figure*}

We show in Fig. \ref{fig: pipeline_images} the imaging results obtained from the crescent test data after different imaging rounds. The image after the second imaging round is shown in the upper right panel, the final image after the fifth imaging round in the lower right panel. The essential image structure is already recovered after the second imaging round (multiscalar imaging with closure properties). This indicates that the multiscalar imaging approach might also be applicable to badly calibrated data and that satisfactory image quality could be achieved even without self-calibration loops. Nevertheless, the use of the amplitudes and full visibility data (imaging rounds 3-5, lower panels) refines the recovered structures and increases coincidence with observed visibilities. Moreover, the steady improvement of the image quality shown in Fig. \ref{fig: pipeline_images} demonstrates that our amplitude conserving hard thresholding approach works in the way intended. We observe a strong contrast between the ring feature and the inner depression (due to sparsity) while the amplitude and total flux is conserved. This would not be available with soft thresholding.

\begin{figure}
    \centering
    \includegraphics[width=0.5 \textwidth]{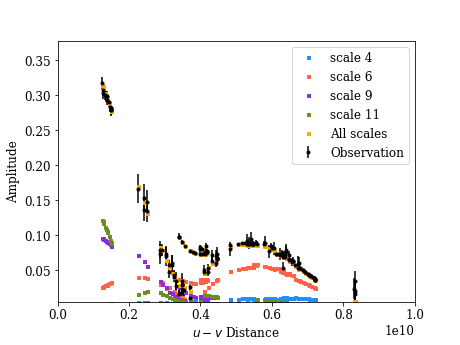}
    \caption{Observed amplitudes (black) and recovered visibilities (yellow) as function of uv-radius for the crescent test data. Moreover, we show the fit of single scales for some selected scales (blue, red, purple, green).}
    \label{fig: crescent_amp}
\end{figure}

We demonstrate in Fig. \ref{fig: crescent_amp} that our final image fits the observed visibilities well. The hard thresholding approach suppresses emission that is not significant for fitting the visibilities, but it does not break the fit to the observed data as soft thresholding would do. In fact, we successfully separated between significant image structures (fitting the visibilities) and noise induced features (very small sidelobe level in the final image).

\begin{figure*}
    \centering
    \includegraphics[width=\textwidth]{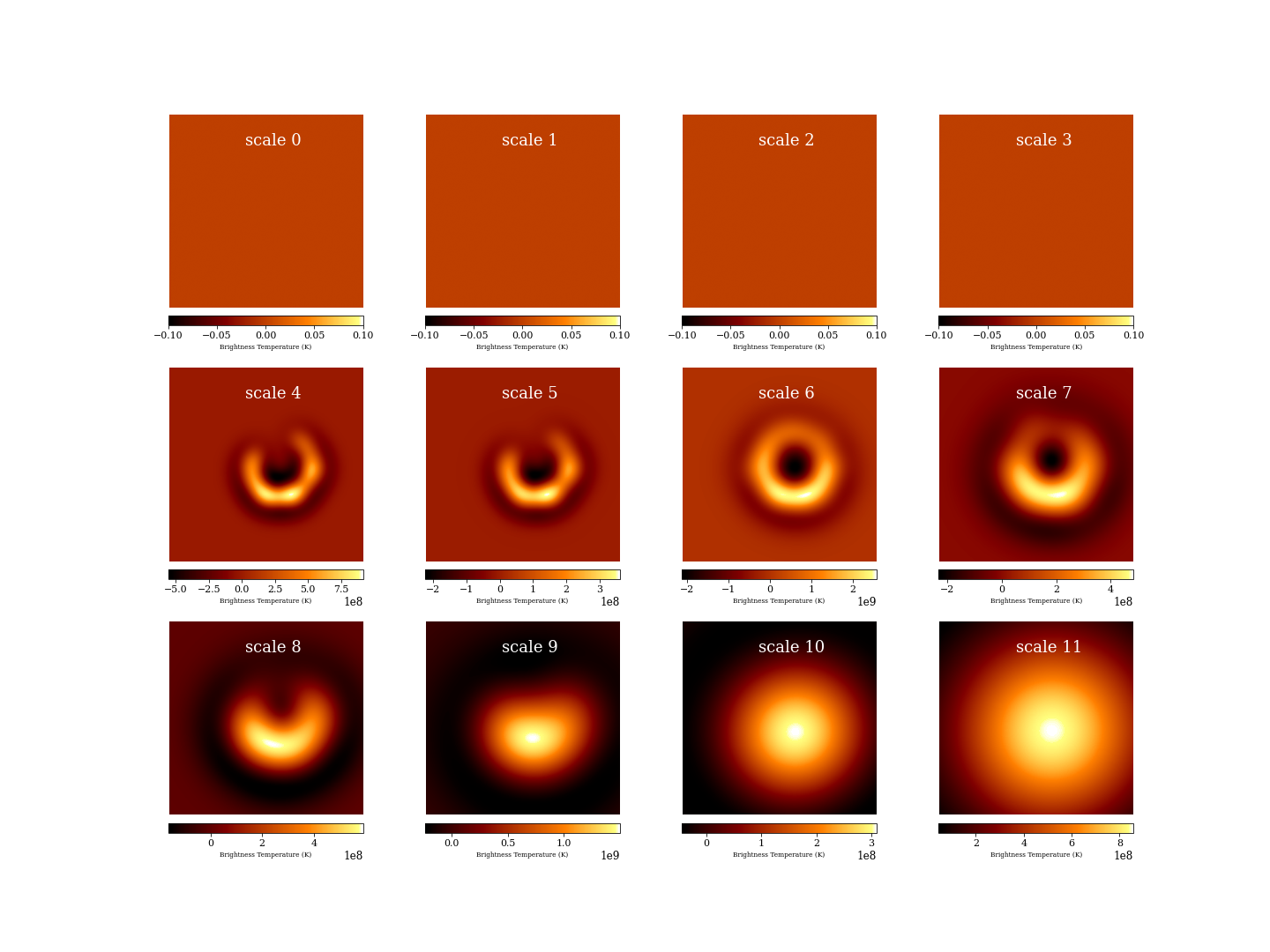}
    \caption{Multiresolution image after imaging round 4. Each panel shows the recovered images from one scale only. The scales are computed with the DoG method with the widths defined in Tab. \ref{tab: scales}. The images is shown for every scale in increasing order from the upper left to the lower right.}
    \label{fig: scalar_images}
\end{figure*}

We present the multiscalar composition of the image in more detail in Fig. \ref{fig: scalar_images}. The panels of Fig. \ref{fig: scalar_images} suggest that different scales are sensitive to different parts of the final image, e.g. an extended Gaussian component (bottom right panel), the ring feature with a central negative peak to compensate for this extended emission (e.g. middle panels) or the asymmetry of the crescent (bottom left panel). The final high resolution and high contrast image is only visible when summing all the single scale images. Additionally, we present in Fig. \ref{fig: crescent_amp} the fit to the data from the single scales only for some selected single scales, i.e. the ones that have the largest signal according to Fig. \ref{fig: scalar_images}. The various scales are in Fourier domain mostly sensitive to varying parts of the uv-coverage, from the short baselines (scales 9 and 11), over the middle baselines (scale 6) to the longest PV-SMA/JCMT baselines (scale 4) as designed. Moreover, Fig. \ref{fig: scalar_images} demonstrates that there are certain scales that are completely suppressed due to the sparsity promoting imaging pipeline (the smallest scales, top panels). Consequently there is no signal at these scales in Fig. \ref{fig: scalar_images}. This is reasonable as these scales are sensitive mainly to fine structures which could only be sampled at baselines longer than the maximum baseline in the data. Moreover, it is noticeable that the scale that is most sensitive to the longest baselines (PV-SMA/JCMT, the fourth scale in Fig. \ref{fig: scalar_images}) is completely suppressed. That, however, does not necessarily mean these data points do not affect the reconstruction anymore. As can be seen in the ring-like masks presented in Fig. \ref{fig: all_scales}, these data points in fact affect all other scales as well (as the Fourier masks are no steep Heaviside functions), but with reduced importance. However, the suppression of this scale could be a hint that further improvement of the method may be available by treating the weighting coefficients $w_j$ in Eq. \eqref{eq: l0_pen} as free parameters.

\subsection{Proof of concept}
\begin{figure*}
    \centering
    \includegraphics[width=\textwidth]{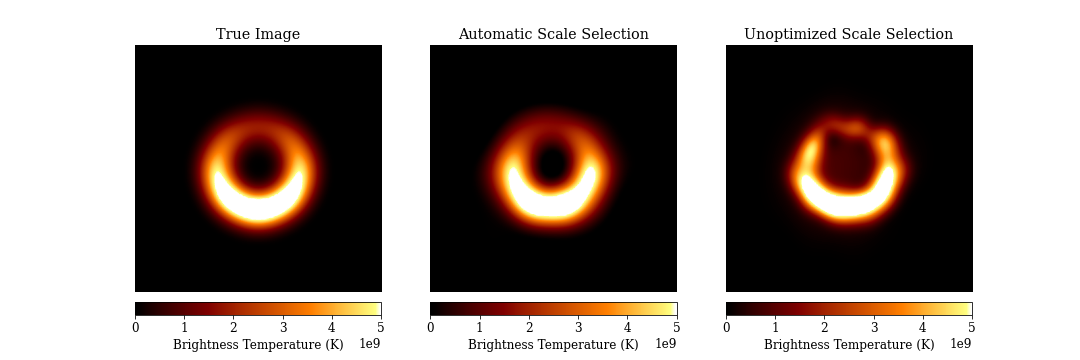}
    \caption{Reconstruction of the crescent image. Left panel: True image. Middle panel: Reconstruction with the selection of scales specified in Tab. \eqref{tab: scales}. Right panel: Reconstruction with scale widths that are a power of two (discrete wavelet transform).}
    \label{fig: proof_of_concept}
\end{figure*}

One of the principal ideas of this paper is to define a flexible wavelet dictionary which adapts smoothly to the uv-coverage. We now prove this concept. We present in Fig. \ref{fig: proof_of_concept} a reconstruction with the complete pipeline with the selection of scales specified in Tab. \ref{tab: scales} and with a coarser grid that would be available for instance with the less flexible a-trou wavelet transform (right panel): $\tilde{\Sigma} = [1, 4, 8, 32]$ (in units of $1.98\,\mu\mathrm{as}$ pixels). We used only every second power of two here for demonstration purposes to enhance the effect of a less fine grid of scales.

The crescent structure is much more robustly recovered with our selection of scales. This is expected, as illustrated by Fig. \ref{fig: scalar_images}. The smaller scales respond to different aspects of the fine structure of the crescent test image, such as the ring like emission, the narrow central ring line or the southern emission peak. The larger scales compress the extended emission. The final high resolution image is only visible by the sum of all these scales. The artificial selection of scales $\tilde{\Sigma}$ has a less complex separation of scales. The complex conglomerate of multiple structure features has to be compressed in only one or two scales. Due to the coarse gridding of widths in $\tilde{\Sigma}$, the algorithm is forced to utilize to small scales which are not able to compensate the bad fitting of the unconstrained minimization. Our automatic scale selection outperforms over this rigid choice of scales because of a more suitable smoothing and thresholding due to adaptive  steps in the scale selection, and hence a more rigorous compression of structure information.

That said, it should be mentioned again that the wavelet dictionaries are complete regardless of the selection of scales. Hence, theoretically the same image can be represented by both wavelet dictionaries regardless of the special choice of scales. The dependency of the reconstruction on the selection of scales is induced by the imaging pipeline (recall that the objective functional is not convex and hence only convergence to a local minimum can be assured). It is easier to recover the image feature at a specific scale, if this scale is well covered by measurements which helps global convergence with our imaging pipeline. On the other hand, a deconvolution at a less well covered scale is more uncertain and possibly fails in the reconstruction of some features.

One may ask now whether progressively refining of the grid of scales should further increase the accuracy of image restoration. while it is principally expected, it also comes with the cost of increased computation time and requires more complexity. In this regard, our automatic scale selection may be viewed as a viable optimum and data-driven approach.

\subsection{Regularization Parameter}
Our algorithm depends on significantly fewer critical parameters that need to be specified by the user. The user only needs to define the regularization parameter $\alpha$ controlling the size of the penalty term, in contrast to the RML methods requiring multiple penalty terms (e.g. with MEM, l1, TV, TSV ... penalty terms) balanced by the term weightings. All other parameters in DoG-HiT are determined automatically from data: the widths of the DoG-dictionary are defined by the automatic procedure described in Sec. \ref{ssec: selection_scales} and the total flux could be identified with the zero-spacing flux which can be measured or estimated. Parameters corresponding to the numerical minimization methods (stepsize, number of iterations, relative tolerance) have only a minor impact on the final result as long as convergence is assured. We present a more quantitative analysis of the impact of the regularization parameter $\alpha$ on the reconstruction in Appendix \ref{app: var_alpha}. In a nutshell, if the regularization parameter is too small, the visibilities are overfitted by a greedy model with a high background level. For higher regularization parameters, the penalty term becomes more important: the background flux level is decreased and the greedy, blobby model becomes more uniform. The best fit is achieved. On the other side, if the regularization parameter is chosen to big, the sparsity penalization dominates the objective functional. The hard thresholding suppresses significant image information and the image is badly fitted with a small number of large wavelet scales.

\section{Comparison to alternative imaging algorithms} \label{sec: comparison}
We compare our image reconstruction with the image reconstructions by standard H\"ogbom CLEAN and the RML method available in the ehtim software package. We utilize the weighting of the data terms for RML reconstructions that was used for \citet{eht2019d} and apply their four-round imaging pipeline published in the EHT data release\footnote{Available under \url{https://github.com/eventhorizontelescope/2019-D01-02}}. The CLEAN reconstructions are performed with the circular window available in the EHT data release \footnote{\url{https://github.com/eventhorizontelescope/2019-D01-02}} and are restored with a $20\,\mu as$ restoring beam. It is worth noting that the RML scripts used for this imaging were extensively optimized for the observations of M87 with the EHT, so excellent reconstructions are expected for this comparison for RML. On the other hand, in contrast to DoG-HiT, those excellent reconstructions required many different parameters to be specified.

\subsection{Qualitative Comparison} \label{ssec: qualitative comparison}
\begin{figure*}
    \centering
    \includegraphics[width=\textwidth]{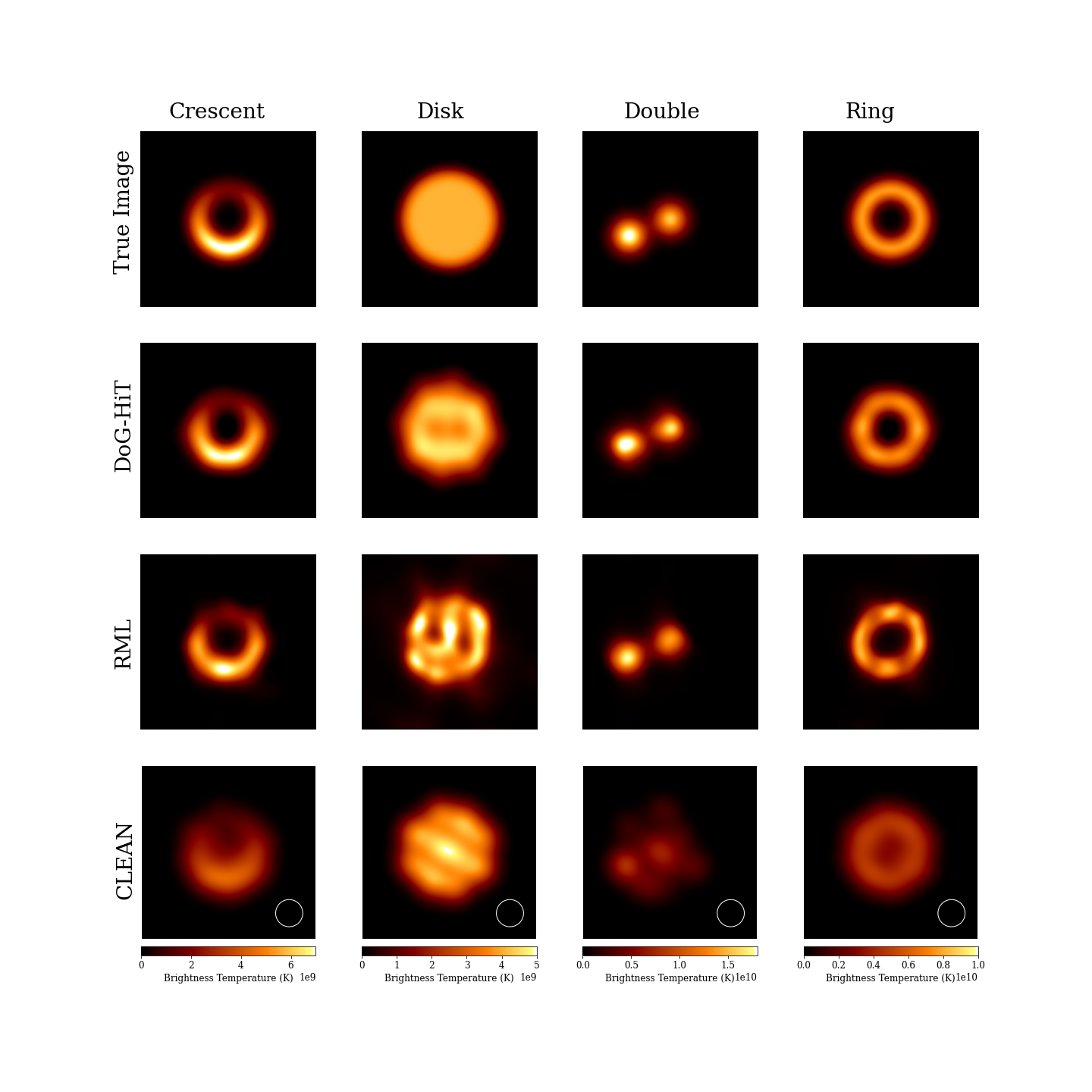}
    \caption{Comparison of the reconstructions with various imaging algorithms. We show in the upper line the true images (crescent, disk, double, ring). In the second to fourth line we present the image reconstructions with DoG-HiT, RML and CLEAN respectively.}
    \label{fig: summary}
\end{figure*}

We show in Fig. \ref{fig: summary} our test image reconstructions on the set of test data presented in Sec. \ref{ssec: testdata}. Our image reconstruction shows a greater resolution than the CLEAN images. Moreover, we achieve a greater contrast between image features and background noise levels than the CLEAN algorithm, i.e. sharper edges in the recovered images. 

Compared to the powerful RML imaging method, our algorithm achieves comparable resolutions. This comes somehow surprising as we probe the observed images with extended basis functions. In particular, we are able to recover some of the fine structure that is not visible in the RML reconstructions. We find the correct crescent-shaped North-South asymmetry in the crescent image, the fine ring ridgeline in the ring image and the correct peak values in the double Gaussian image. Moreover, we find a greater contrast between the ring-like features in the ring and crescent images and the central depression, compared to that observed in RML image. However, the inner 'no-emission' radius is smaller than in the true images with DoG-HiT while the spherical shape remains better recovered. This region is significantly better recovered by the RML algorithm. Moreover, RML appears to perform better in resolving the ring and crescent features transversely. 

Notably, our algorithm also succeeds in the reconstruction of smooth extended emission, e.g. of the disk image. The reconstruction of the disk is quite accurate and comparable to the reconstruction with CLEAN. It does not manifest the greedy image disk features or background emission present in the RML reconstruction. The ring image demonstrates that DoG-HiT is able to fit uniform emission (ring extension) and sharp features (ring edges) simultaneously. The CLEAN reconstruction  of the ring lacks the proper reconstruction of the sharp ring edges and the central depression. The RML reconstruction fits the central depression well, but the ring brightness distribution is less homogeneous than in the DoG-HiT reconstruction. In this way DoG-HiT combines the major advantages of RML reconstructions (super-resolving structures) and CLEAN (high dynamic range sensitivity to extended structures), at the same time also reducing the drawbacks of either of these two methods. It should therefore be well suited for imaging problems arising in the context of EHT observations in which the demand on recovery of information contained on smallest accessible scales requires simultaneous robust imaging of extended structures (jet). Performance of Dog-HiT under these conditions will be discussed in in Sec. \ref{sec: physical} using simulated data with a wide range of spatial scales.

\subsection{Quantitative Comparison}
We now compare the various imaging algorithms in a more quantitative way using a measure of the their relative error, 
\begin{align}
    \mathrm{err} = \frac{\norm{\mathrm{recovered\:solution} - \mathrm{exact\:image}}}{\norm{\mathrm{exact\:image}}}. \label{eq: rel_error}
\end{align}
We present the relative errors of the reconstructions in Tab. \ref{tab: accuracy}. The comparison may be somewhat unfair for CLEAN given the large beam size compared to the size of the structures, but a final convolution with a synthetic point spread function is the common standard in radio astronomy. We present the relative error of the reconstruction both without blurring (as it is standard for RML and DoG-HiT) and with blurring by 1/2 of the beam size and the full beam size (as it is standard for CLEAN). The super-resolving DoG-HiT reconstructions are getting worse with larger restoring beam, while for CLEAN the opposite is true. DoG-HiT tops the challenge for three of the four test images (crescent, disk, ring) and performs similar to RML for narrow structures (crescent, double). Overall, we can conclude that DoG-HiT is able to achieve a similar precision as current imaging algorithms, but alleviates some of the limitations of both CLEAN (no superresolution) and RML methods (sensitivity to smooth extended features).

\begin{figure*}
    \centering
    \includegraphics[width=\textwidth]{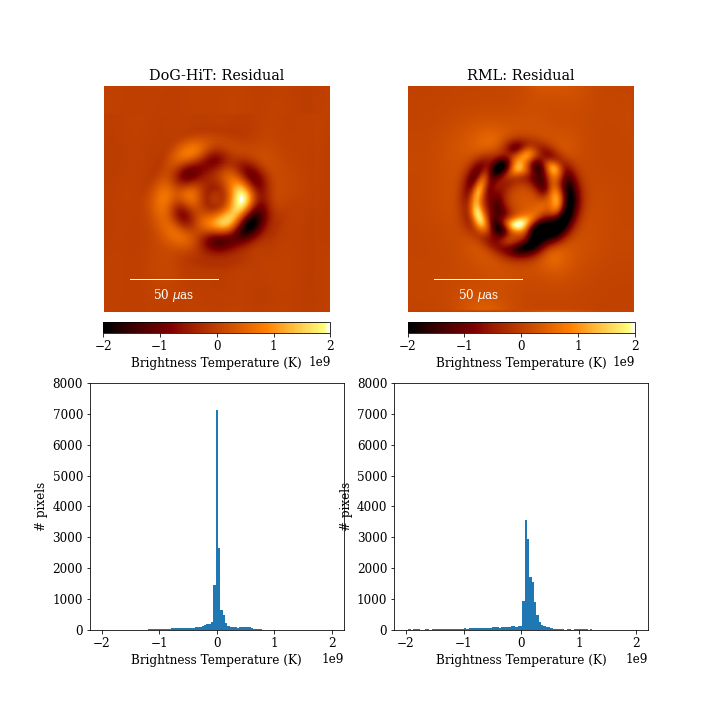}
    \caption{Residuals of the reconstructions with DoG-HiT (left) and RML (right) for the ring image. Upper panels: True image subtracted from the reconstructed image. Lower panels: Histogram of the residual distribution.}
    \label{fig: residual}
\end{figure*}

We present in Fig. \ref{fig: residual} the residuals of the reconstructions of the ring feature with RML and with DoG-HiT. The residuals for both imaging methods are ring-shaped and spatially correlated, indicating that there is still unrecovered structure. However, the histograms of the residuals in the lower panels  of Fig.~\ref{fig: residual} demonstrate overall a very good reconstruction. The pixel residual distribution is well approximated by a narrow Gaussian distribution in both cases. Nevertheless, the residual distribution for DoG-HiT is slightly more narrow and less skewed, which agress well with the overall slightly smaller relative error listed in Table~\ref{tab: accuracy}.

\begin{table}[]
    \centering
    \begin{tabular}{l l| c c c c}
    Blurring & & Crescent & Disk & Double & Ring\\ \hline
         &DoG-HiT & 0.156& 0.138& 0.167& 0.139\\
    $0\,\mathrm{\mu as}$ &RML & 0.16& 0.266& 0.164& 0.211\\
         &CLEAN & 1.121& 1.282& 1.427& 1.082\\ \hline
         &DoG-HiT & 0.219& 0.144& 0.191& 0.215\\
    $10\,\mathrm{\mu as}$ &RML & 0.238& 0.219& 0.234& 0.245\\
         &CLEAN & 0.294& 0.568& 0.658& 0.285\\ \hline
         &DoG-HiT & 0.414& 0.203& 0.402& 0.411\\
    $20\,\mathrm{\mu as}$ &RML & 0.433& 0.275& 0.443& 0.441\\
         &CLEAN & 0.399& 0.156& 0.556& 0.396\\ \hline
    \end{tabular}
    \caption{Relative errors of the reconstructions shown in Fig. \ref{fig: summary}.}
    \label{tab: accuracy}
\end{table}

\subsection{Transverse Resolution}
\begin{figure}
    \centering
    \includegraphics[width = 0.5 \textwidth]{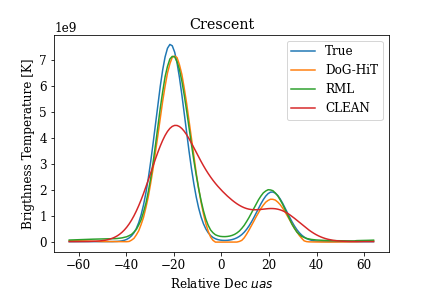}
    \caption{Profiles of the recovered crescent images in North-South direction and central right ascension.}
    \label{fig: profile}
\end{figure}

We study the transverse resolution of the algorithms with the crescent image in this section. We present in Fig. \ref{fig: profile} the profiles of the true (blue) and the recovered crescent images in North-South direction at central right ascension. We recover the correct double peak structure with North-South asymmetry both with the RML method and with DoG-HiT. CLEAN is not able to reproduce this fine structure sufficiently. Regarding transverse resolution of the ring features and the central depression, RML and DoG-HiT perform equally well, recovering approximately the correct widths of the Gaussian blurred ring and the correct depth of the central depression. However, DoG-HiT recovers a zero-flux central depression which is not captured in the true image. We computed the blurring beam size that maximizes the correlation between the (blurred) true image and the recovered images to quantify the resolutions. The largest correlation between the DoG-HiT reconstruction and the true image was achieved if the true image is blurred by a beam with widths $6.1\,\mu\mathrm{as}$. That means that DoG-HiT was able to reproduce image features down to a resolution of approximately $6\,\mu\mathrm{as}$. For the RML reconstruction we found a maximal correlation for a beam of $5.2\,\mu\mathrm{as}$ similar to DoG-HiT. This resolution is expected due to the reverse taper of $5\,\mu\mathrm{as}$ applied in the ehtim pipeline. For CLEAN we found a widths of $19.7\,\mu\mathrm{as}$ coinciding well with the applied point spread function of the array.

\subsection{Simplicity and Performance}
The five imaging steps presented in Sec. \ref{ssec: pipeline} may not appear as a simple approach to imaging. However, one should recall that this strategy resembles typical steps in the imaging of interferometric data with CLEAN: imaging in several loops of cleaning and self calibration. Hence, our lengthy pipeline is not more complex than automatic cleaning scripts. 

More importantly, DoG-HiT only takes into account a very limited number of regularization parameters, namely only the prior guess for the total flux and the biasing parameter $\alpha$ which controls the weight of the hard thresholding regularization term (see Eq. \eqref{eq: second_round_problem}). Apart from these parameters, only solver related choices such as stepsize or relative tolerances need to be specified. This is a first step towards a more unsupervised imaging algorithm in which crucial choices for the imaging procedure (i.e. selection of window or regularization parameters) come directly from data and are not manually selected doing the analysis.

The down side of this simplicity is that DoG-HiT is also less flexible compared to RML methods. RML methods combine different type of penalizations and prior assumptions that could be relatively weighted according to specifications of a certain data set. While it is a promising news that wavelet sparsity promoting algorithms are similar or, in some settings, even better performing than RML methods, this conclusion cannot be automatically generalized. In particular, more advanced calibration issues could add an additional layer of complexity to the problem. Furthermore, the hard thresholding method used in DoG-HiT may limit the dynamic range of the reconstruction, i.e. the minimal flux that can be recovered. A more rigorous study of this drawback should be made in subsequent works and applications.

Furthermore, it is a serious disadvantage of our algorithm that it presently requires considerably more time and computing resources than the fast RML methods, because the image is overcompletely represented by wavelet scales. 

\section{Physical source model} \label{sec: physical}
\begin{figure*}
    \centering
    \includegraphics{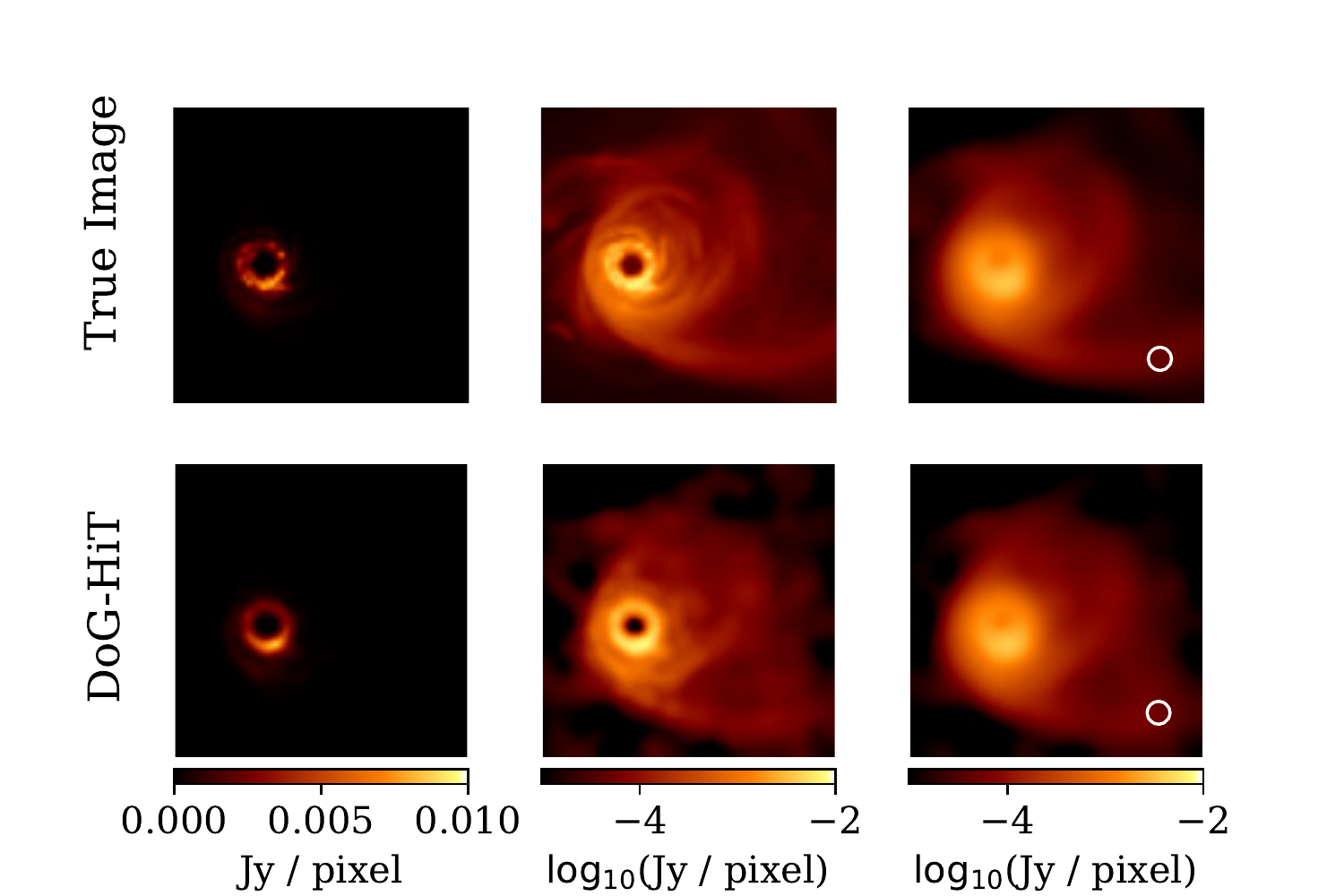}
    \caption{Reconstruction of synthetic M87 observation with a possible ngEHT array taken from the ngEHT Analysis Challenge \citep{ngehtchallenge}. The true image is presented in the upper panels, the reconstruction with DoG-HiT in the lower panels. The left panels show the ground truth and the recovered images in linear scale, the middle panels in logarithmic scale (i.e. highlighting the extended emission from the jet basis) and the right panels compare the ground truth an the recovered image both smoothed with a restoring beam of $20\mu as$.}
    \label{fig: howes_summary}
\end{figure*}

\begin{figure}
    \centering
    \includegraphics[width=0.5\textwidth]{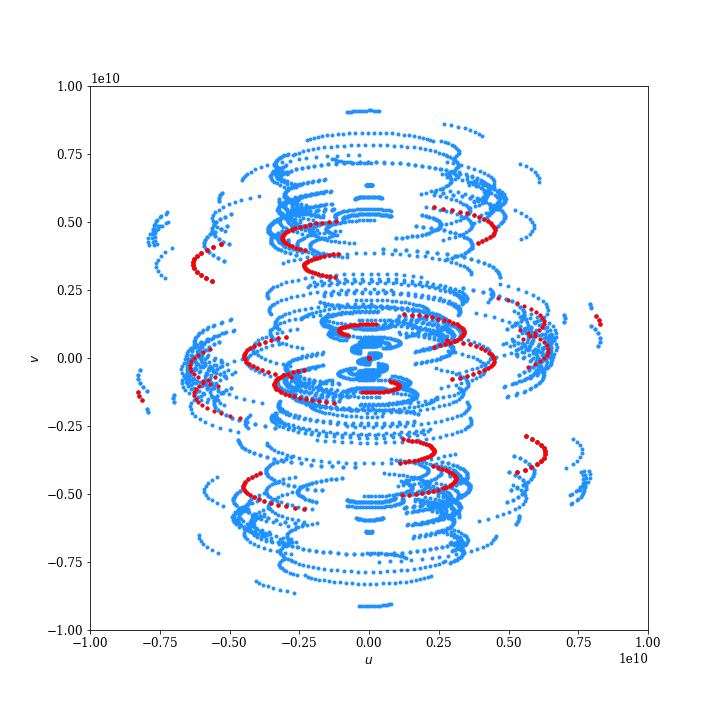}
    \caption{uv-coverage of a synthetic ngEHT observation of M87 at $230\,\mathrm{GHz}$. The uv coverage with the EHT 2017 antennas only is plotted in red. For more details see \citet{ngehtchallenge}.}
    \label{fig: uv-coverage}
\end{figure}

To demonstrate the performance of the algorithm on structures covering a wider range of spatial scales, we present here a DoG-HiT image reconstruction made from synthetic data from the first ngEHT Analysis Challenge \footnote{\label{fn: ngeht} Available under \url{https://challenge.ngeht.org/challenge1/}}, which emulate the black hole shadow and the jet base in M\,87 as observed with a possible ngEHT configuration \citep{ngehtchallenge}. The ngEHT is a planned, but not finally proposed future global VLBI array designed to produce real time movies of the dynamics in the extreme vicinity of a black hole and the innermost jet region \citep{Doeleman2019}. The ngEHT builds up on the enormous success of the EHT and will extend the EHT science with higher dynamic ranges, sensitivity and resolution. It is believed to deliver novel groundbreaking results for the formation of jets, accretion physics and general relativity tests \citep{Doeleman2019}. In particular the dense uv-coverage (including short baselines) and high sensitivity of the ngEHT as compared to the current EHT allow for the reconstruction of the extended jet emission. The reconstruction and the true image are compared in Fig.~\ref{fig: howes_summary} in linear (left column) and logarithmic (middle and right column) scales, with the latter employed for highlighting the extended emission. The simulated source structure is taken from a MAD GRMHD simulation of a rapid spinning black hole surrounded by an accretion disk with electron heating from reconnection \citep{ngehtchallenge, Mizuno2021, Fromm2022}. The simulated visibilities are calculated for a \textit{possible} template ngEHT configuration at $230\,\mathrm{GHz}$ that is used throughout the ngEHT Analysis Challenge \citep{ngehtchallenge} and might be realized in the final concept of the array. It contains the eleven current EHT sites (ALMA, APEX, GLT, IRAM-30 m, JCMT, KP, LMT, NOEMA, SMA, SMT, SPT) and ten additional stations from the list of \citet{Raymond2021} (BAR, OVRO, BAJA, NZ, SGO, CAT, GARS, HAY, CNI, GAM). HAY, OVRO, and GAM are 37, 10.4, and 15 m antennas respectively. All of the remaining additional antennas are assumed to be of 6\,m in diameter. The synthetic visibilities are simulated with a 10 sec averaging time and with alternating 10 min observation scans and 10 min gaps. The resulting uv-coverage is presented in Fig.~\ref{fig: uv-coverage}. For more information on the generation of the ground truth image and the synthetic observation we refer the interested reader to the description of the ngEHT Analysis Challenge available at the link listed in footnote~\ref{fn: ngeht}.

The DoG-HiT reconstruction in Fig.~\ref{fig: howes_summary} represents accurately the central ring-like structure. This is expected, judging from the successful reconstructions obtained in Sec.~\ref{ssec: qualitative comparison} on the compact crescent images from a much sparser synthetic observation. In addition to this, Fig.~\ref{fig: howes_summary} demonstrates that DoG-HiT also reproduces very well the extended emission from the jet base (middle and right panels). The structural details of the jet base are compressed on much larger scales than the smaller ring feature and are less bright (hence only visible in logarithmic scale). This result demonstrates the ability of DoG-HiT to work on images with a wide range of spatial scales and large dynamic range.

\section{Conclusion} \label{sec: conclusion}
In this paper, we presented a novel interferometric imaging algorithm which is capable of adapting to the Fourier domain coverage of observations and particularly applicable to sparse uv-coverages. Our imaging algorithm models the image as a sum of difference of Gaussians wavelet functions. This wavelet dictionary is more flexible than the usual discrete a-trou wavelet transform and allows us to select the scales to adapt to the uv-coverage.

We formulate the imaging problem as an optimization problem with an objective functional consisting of the reduced $\chi^2$ of the recovered closure properties (closure phase and logarithmic closure amplitudes) and an $l_0$-pseudonorm sparsity term in the wavelet domain. As this objective functional is still invariant against rescaling of the image guess, we also add a total flux constraint. The resulting objective functional is non-smooth and non-convex, but could be solved by an iterative hard thresholding splitting algorithm for which local convergence to a steady point is known. Due to non-convexity, global convergence cannot be assured, but practice shows that local minima could be avoided by proper initial guesses. Our algorithm is amplitude- and total flux-conserving, in contrast to schemes using soft thresholding. Together with a more thorough separation of image features and sidelobes by a flexible wavelet dictionary analysis, this is expected to bring significant improvements in imaging of VLBI data with strongly varying and scale-dependent noise.

We present a complete imaging pipeline ready for application. Our imaging pipeline consists of five imaging rounds, where we refine the initial imaging results from the closure properties in an iterative imaging/self-calibration loop which uses the amplitude and phase information. We apply for the first time a multiresolution constraint for these refinement steps. Moreover, we prove stability of our pipeline in practice on synthetic data.

Comparisons of imaging performance on the synthetic data show that DoG-HiT achieves super-resolution and outperforms CLEAN in the reconstruction of fine structure (super-resolving) and that it comparable to RML methods in terms of accuracy of reconstruction. DoG-HiT succeeds in the reconstruction of smooth extended emission components, where it outperforms RML. It effectively combines the strengths of CLEAN and RML methods and reduces their specific weaknesses. DoG-HiT should therefore be well suited for application to targets with a wide range of spatial scales, for which it may be outperforming current RML reconstructions in the context of better recovering smoother emission on large scales. We have demonstrated this capability on a synthetic data set from the first ngEHT challenge, with excellent reconstructions achieved for of both the small scale inner ring-like structure and the faint, larger scale emission from the jet base. It should also be noted that the DoG-HiT reconstruction accurately reproduces features with a strong contrast between emission and the background. At the same time, DoG-HiT presently introduces some systematic inaccuracies (e.g. a limited dynamic range) into the reconstruction, and this needs to be addressed in future works.

\section*{Software Availability}
We will make our imaging pipeline and our software available soon in a suitable way. Our software makes use of the publicly available ehtim \citep{Chael2018}, regpy \citep{regpy} and WISE software packages \citep{Mertens2015}.

\section*{Acknowledgements}
We thank F. Roelofs, C. Fromm, L. Blackburn, G. Lindahl, A. Raymond, S. Doeleman and the team of the ngEHT Analysis Challenge for providing their data set and for useful discussions. HM received financial support for this research from the International Max Planck Research School (IMPRS) for Astronomy and Astrophysics at the Universities of Bonn and Cologne.

\bibliography{lib}{}
\bibliographystyle{aa}


\appendix

\section{Fixed point property of proximity operators} \label{sec: fixed_point}
Let $\hat{x} \in \mathrm{argmin}_s H(x)$ and let $x \in \mathbb{X}$. Then it is: 
\begin{align}
    H(x) + \frac{1}{2\tau} \lVert s - \hat{s} \rVert_\mathbb{X} \geq H(\hat{s}) + \frac{1}{2\tau} \lVert \hat{s} - \hat{s} \rVert_\mathbb{X},
\end{align}
as $\hat{s}$ is in the argmin of $H$ and $\tau \geq 0$. Vice versa, let $\hat{s}$ be the solution to $\hat{s} = \mathrm{prox}_{\tau, H}(\hat{s})$, then it follows, see Eq. \eqref{eq: prox2}: 
\begin{align}
    0 = \hat{s} - \hat{s} \in \tau \partial H [\hat{s}],
\end{align}
which suffices to show for a convex, proper and lower semicontinuous functional that $\hat{x}$ is in the argmin of $H$.

\section{Variation of regularization parameter} \label{app: var_alpha}
\begin{figure*}
    \centering
    \includegraphics[width=\textwidth]{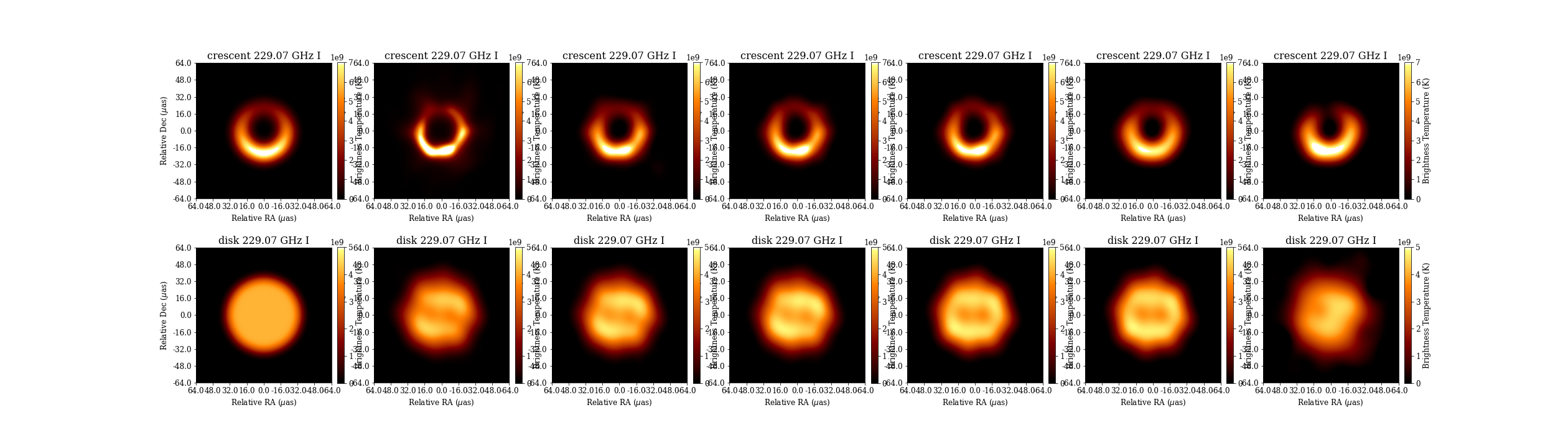}
    \caption{Reconstructions with varying regularization parameter $\alpha$. Most left panels: True images. Middle panels from left to right: $\alpha \in \{0, 10^{-3}, 10^{-2}, 10^{-1}, 10^{0}, 10^{1}\}$.}
    \label{fig: var_alpha}
\end{figure*}

We discuss in this subsection the impact of the regularization parameter $\alpha$. We show in Fig. \ref{fig: var_alpha} the reconstruction of the crescent image and of the disk image with varying regularization parameter $\alpha$. The most left panels show the true image, the second left panels the unconstrained reconstructions, e.g. $\alpha = 0$. The middle panels show from left to right the reconstruction results obtained with increasing values of $\alpha$. We present in Tab. \ref{tab: var_alpha} the relative precisions \eqref{eq: rel_error} of the different reconstructions. The reconstructions are worse for too small $\alpha$ and too big $\alpha$. The best fit value lies somewhere in between.

The reconstructions for very small $\alpha$ show a greedy and too-fine resolving model that differs significantly from the true image. Moreover, fainter sidelobes and background emission is visible in the reconstruction. These models overfit the observed visibilities, i.e. the observed visibilities are fitted exactly, but the gaps in the uv-coverage are filled by high oscillating fits.

For intermediate $\alpha$ the reconstruction is best. The unconstrained reconstruction is modeled with few (due to sparsity) extended, smooth wavelet functions. This approach effectively smoothes the fit to the visibilities and fills the gaps in the uv-coverage with smooth fits. In this spirit the sparsity approach in the wavelet basis has a similar effect on the data such as total variation, total squared variation and maximum entropy penalizations. Moreover, sidelobes in the image are suppressed by the hard thresholding.

On the other end of the table the reconstructions worsen again for too large reconstruction parameters. In these cases the penalty term dominates the objective functional in the forward-backward minimization. The image is modelled with too few wavelet scales. The result is a blurry reconstruction. Moreover, the hard thresholding minimization procedure cuts significant fainter features, e.g. the northern emission in the crescent test image (upper most right panel in Fig. \ref{fig: var_alpha}).

\begin{table}[]
    \centering
    \begin{tabular}{c | c c c c c c}
        $\alpha$ & $0$ & $10^{-3}$ & $10^{-2}$ & $10^{-1}$& $10^{0}$& $10^{1}$ \\
         crescent & $0.345$ & $0.17$ & $0.148$ & $0.148$ & $0.169$ & $0.254$\\
         disk & $0.202$ & $0.164$ & $0.154$ & $0.137$ & $0.138$ & $0.231$
    \end{tabular}
    \caption{Relative error of the DoG-HiT reconstruction with varying assumptions on the regularization parameter $\alpha$.}
    \label{tab: var_alpha}
\end{table}

\end{document}